# Room-temperature quantum transport signatures in graphene/LaAlO$_3$/SrTiO$_3$ heterostructures

Giriraj Jnawali,[1,2] Mengchen Huang,[1,2] Jen-Feng Hsu,[1,2] Hyungwoo Lee,[3] Jung-Woo Lee,[3] Patrick Irvin,[1,2] Chang-Beom Eom,[3] Brian D'Urso,[1,2] and Jeremy Levy[1,2*]

[1]Department of Physics and Astronomy, University of Pittsburgh, Pittsburgh 15260, USA

[2]Pittsburgh Quantum Institute, Pittsburgh, PA 15260, USA

[3]Department of Materials Science and Engineering, University of Wisconsin-Madison, Madison, WI 53706, USA

[*]Corresponding author: jlevy@pitt.edu

Weak localization (WL) and weak antilocalization (WAL) are quantum interference effects[1-4] resulting from electron phase coherence and spin-orbit interactions in 2-dimensional (2D) electron systems. WL results from constructive interference between pairs of time-reversed closed-loop electron trajectories and provides a positive correction to the Drude resistivity.[2, 4] Spin-orbit coupling (SOC) leads to suppressed backscattering due to destructive interference, leading to WAL and a negative correction to the Drude resistivity.[3]

The intrinsic SOC of graphene is weak;[5] however, charge carriers in graphene possess a pseudospin degree of freedom, which arises from the degeneracy introduced by the two inequivalent atomic sites per unit cell in the graphene honeycomb lattice being comprised of A and B sublattices [**Figure 1**(a) and (b)].[6, 7] Recent discovery of unique quantum transport phenomena in graphene, such as unconventional half-integer quantum Hall effect[7, 8] and Klein tunneling[9-11] is a direct consequence of non-trivial Berry phase of $\pi$ induced by pseudospin rotation. Pseudospin in graphene can be utilized to store and manipulate information, which is analogous to the spin degree of freedom in spintronics.[12-16] Because of pseudospin rotation, each scattering process has a phase difference of $\pi$ between two time reversal pair in a closed quantum diffusive path. This results in destructive interference that suppresses backscattering, leading to WAL in graphene, which is analogous to the role of SOC (**Figure 1**(c) and (d)) in ordinary semiconductors. WAL is theoretically expected in graphene in the absence of inter-valley and chirality breaking scattering.[17] Most studies have not presented clear evidence of WAL via negative magnetoconductance,[18-22] most likely due to presence of point defects in graphene samples that locally break the sublattice degeneracy and smooth out $\pi$-phase contribution. Experimental signatures of WAL observed in high-quality epitaxial graphene samples are attributed to suppressed point defects.[23] Interestingly, a WL to WAL transition is achieved in high quality exfoliated samples[24] by decreasing the ratio of the dephasing length to the symmetry-breaking length via decoherence and carrier-density control.

Preservation of pseudospin quantum interference up to room temperature (RT) is of fundamental importance for a variety of proposed applications,[12-16, 25, 26] but thermal perturbations typically suppress the phase coherence and hinder practical use of the devices. RT operation can be achieved if the high graphene mobility is consistently maintained over a broad temperature range and phonon-scattering contribution is significantly reduced. The mobility of graphene field-effect devices fabricated on $SiO_2$/Si substrates is typically reduced at RT due to scattering with surface polar modes.[20, 27, 28] A significant improvement in quality was achieved by fabricating graphene devices on hexagonal-boron nitride (hBN) due to the extremely flat surface and strong in-plane bond of hBN.[29] Similarly, high quality graphene-field-effect devices were realized using high-$k$ ferroelectric substrates,[30] which also resulted in novel graphene-based non-volatile memory devices. To our knowledge, WAL is not observed at RT in these hybrid structures.

Integrating graphene with various functional materials provides opportunities to develop multifunctional devices as well as to explore intrinsic graphene quantum transport phenomena. Among multi-functional materials, the complex-oxide heterostructure $LaAlO_3$/$SrTiO_3$ has attracted great interest because of the intriguing physics discovered at the interface[31] such as two-dimensional conductivity,[32] superconductivity,[33, 34] magnetism,[35-38] and spin-orbit coupling.[39, 40] One distinctive property of this system is the existence of a sharp metal-insulator transition (MIT) at a critical thickness of $LaAlO_3$, $d_c \sim 4$ unit cell (uc), at or above which the interface becomes conducting.[32] Just below this thickness, $d \sim 3$ uc, the $LaAlO_3$/$SrTiO_3$ interface can be locally and reversibly switched between conductive and insulating phases using conductive atomic force microscope (c-AFM) lithography,[41, 42] allowing the fabrication of reconfigurable nanoscale electronic devices. It has recently been demonstrated that this technique can also be used to create devices on graphene/$LaAlO_3$/$SrTiO_3$ heterostructures.[43] Since the coexistence of multiple phases in oxide heterostructures is largely associated with a manifold of electronic degrees of freedom and strong correlations between them, unique interactions with the nearby graphene layer may be expected.

Here we report fabrication and magnetotransport characterization of high mobility graphene field-effect devices on a complex-oxide heterostructure, LaAlO$_3$/SrTiO$_3$. These devices exhibit quantum transport signatures such as anomalous quantum Hall effect and WAL behavior over a broad temperature range. Remarkably, the mobility is only weakly dependent on temperature (up to 300 K), suggesting weak coupling with substrate optical modes and suppressed electron-phonon scattering contributions. The unique interaction with the oxide heterostructure is associated with direct observation of pseudospin chirality via WAL up to RT in CVD-grown graphene. We investigate WAL via magnetotransport measurements as a function of carrier density and temperature. We will restrict our focus to transport in the graphene, which is electrically isolated from the LaAlO$_3$/SrTiO$_3$ interface.

Graphene samples used in this work are synthesized using atmospheric pressure chemical vapor deposition (APCVD) growth method[44] on ultra-flat Cu wafers[45] and subsequently transferred onto pre-patterned LaAlO$_3$/SrTiO$_3$ substrates. Following transfer, deep-UV lithography and oxygen plasma cleaning processes selectively remove unwanted graphene.[46] Graphene samples are patterned into Hall bars with nominal channel lengths of $L = 3$ μm and widths of $W = 5$ μm by anodic etching using c-AFM.[47, 48] Details about preparation of LaAlO$_3$/SrTiO$_3$ heterostructure, graphene growth, transfer, and *in-situ* c-AFM Hall bar patterning are described in the Methods section and in the Supporting Information. Typical graphene device layout and transport measurement scheme are shown in **Figure 2**(a). Four-terminal longitudinal and transverse resistances are measured simultaneously through top-gated electrodes using low frequency (1 − 10 Hz) lock-in detection. Other nearby electrodes are designed to be in contact with the LaAlO$_3$/SrTiO$_3$ interface, and are electrically isolated from the graphene and top-gated electrodes. The carrier density is tuned by a "side gate" voltage $V_{sg}$ applied through these electrodes, which acts from the same plane of the graphene device. Alternatively, we can use a "back gate" voltage $V_{bg}$, which is a gate that is applied to the bottom of the SrTiO$_3$ substrate. Since gating from the back requires larger gate voltages as compared to the side gate voltages, most experiments reported here employ $V_{sg}$ to tune the graphene Fermi level. However, all of the main features have been reproduced for both gating methods (see Supporting Information).

The device exhibits standard ambipolar characteristics, with resistivity modulation ratio more than five at 300 K and twenty at 2 K. **Figure 2**(b) shows the gate dependence of longitudinal resistivity $\rho_{xx}$ of the device at 2 K and 300 K. We observe hysteresis in the voltage at which the Dirac point is reached. The voltage hysteresis at the Dirac point varies with sweeping parameters such as sweep range and rate, but is not attributed to instrumental effects (e.g., lock-in time constant). In general, the hysteresis increases as the sweeping range and the rate increase. Unlike previous reports of hysteresis observed in different graphene devices,[49-53] origin of hysteresis in our device is not clear. A more detailed discussion of the hysteresis and possible causes are described in Supporting Information.

In a magnetic field applied perpendicular to the graphene basal plane, the device exhibits characteristic oscillations in longitudinal resistivity $\rho_{xx}$ and quantization of Hall conductance $\sigma_{xy}$ (**Figure 2**(c)). The quantization is described by $\sigma_{xy} = \pm \nu(e^2/h)$, where $\nu = \pm 4(N + 1/2)$ is the filling factor (number of filled Landau levels (LLs)), $N$ is a non-negative integer, $e$ is the elementary charge, $h$ is the Plank constant, and $\pm$ stands for electrons and holes, respectively. Observation of half-integer quantized Hall plateaus accompanied by vanishing $\rho_{xx}$ is a clear signature of the quantum Hall effect (QHE) in single layer graphene devices.[6, 7] The quantization appears for both sweeping directions with identical QH plateau (Supporting Information). For clarity, only backward sweeping data are shown in **Figure 2**(c) and (d).

We now characterize the electronic properties of our graphene devices by measuring the magnetic field-dependent sheet resistivity $\rho_{xx}$ and transverse resistance $R_{xy}$. At low magnetic fields the linear slope $s = dR_{xy}/dB$ determines the carrier density $n = 1/(s \cdot e)$. The Hall mobility $\mu$ is calculated through the expression $\mu = 1/(n \cdot e \cdot R_{xy})$. **Figure 2**(d) shows the extracted carrier density $n$ and the Hall mobility $\mu$ measured at different gate voltages $V_{sg}$. The carrier density $n$ has a linear dependence with $V_{sg}$ towards both electron and hole sides of the Dirac point (red dashed lines), demonstrating efficient field-effect gating behavior through the complex oxide heterostructure. The density saturates near the Dirac point to a minimum density of $n_{min} \sim 4\times10^{10}$ cm$^{-2}$, which is also apparent from the crossing point of linear fits that lies below zero at $V_{sg} = V_{Dirac} \sim 5$ V. Below $n_{min}$ the Hall resistance becomes significantly non-linear due to the inhomogeneous landscape of carriers in electron-hole puddles[54] and therefore determination of carrier density from the Hall coefficient becomes non-trivial. The relatively low saturation density $n_{min}$ indicates low disorder in our samples. The gate voltage dependence of Hall mobility is also interesting, which shows sharp peak near the Dirac point and slowly approaches towards saturation at either sides of electron and holes around Dirac point. The Hall mobility approximately follows an inverse relation with carrier density, i.e., $\mu \propto 1/n$ (left inset of **Figure 2**(d)). Therefore, based on previous studies,[28, 55, 56] we anticipate charged impurity scattering limits the mobility of our samples, which is improved around low-density regime due to reduced Coulomb interaction of carriers with charged impurities and scattering between them. Such a sharp profile of density dependent mobility was also reported on number of graphene devices supported on different high-$k$ materials and attributed to a direct consequence of strong dielectric screening of charged impurity scattering.[51, 57, 58] The mobility of our device varies from $\sim 1 - 4\times10^4$ cm$^2$/V·s over the entire gate voltage region, which is comparable with the best CVD-grown graphene devices reported to date.[59, 60] Since the Hall mobility depends strongly on the carrier density, it is natural to examine the density-independent mobility $\mu_c$, which excludes the effects of various other sources of disorders. Therefore, we estimate the density-independent mobility $\mu_c$, which is associated with long-range Coulomb scattering, by fitting the sheet conductivity $\sigma_{xx}$ data using the commonly employed self-consistent diffusive transport model.[61] As shown in the right inset of **Figure 2**(d), the model fits the data well with $\mu_c \sim 2\times10^4$ cm$^2$/V·s, further confirming the high quality of these samples.

Now we examine the temperature dependence of transport properties by measuring the sheet resistivity $\rho_{xx}$ and the Hall resistance $R_{xy}$ simultaneously at zero and a fixed field of $B = 5$ T, during warming up the sample. The results are plotted in **Figure 3**(a). Surprisingly, the zero-field sheet resistivity $\rho_{xx}$ rapidly increases from ~350 Ω as the sample is warmed from 2 K, peaks at ~3.5 kΩ near 130 K, and then slowly decreases to ~1.5 kΩ at 300 K. This behavior has been reproduced for both warm-up and cool-down cycles, with minor shift that may be attributed to slight differences in actual sample temperature during each thermal cycle. At $B = 5$ T, the sheet resistivity $\rho_{xx}$ exhibits pronounced oscillations, and the Hall resistance $R_{xy}$ shows well-defined plateaus at each oscillation minimum up to 100 K (indicated by red arrows). Above 100 K, the slope of the Hall resistance changes sign from negative to positive, showing change of polarity from hole-doped at low temperature to electron-doped at high temperatures. The oscillation minima and the Hall quantization matches with the LL filling factors for single-layer graphene, revealing the half-integer QHE in graphene measured as a function of temperature. These features, however, are smeared out above 100 K (towards the electron-doped side), which is caused by thermal broadening of LLs. Interestingly; the shift of Dirac point could also be tuned when the sample is measured with fixed gate. This has been demonstrated by measuring the temperature dependent magnetoresistance with a gate bias of $V_{sg} = 12$ V (see Supporting Information), which shows the shift of Dirac point towards low temperature side due to positive gating. Due to the shift of Dirac point towards low-temperature, i.e., at ~35 K, well-developed QH plateaus and pronounced SdH oscillations are visible for both electrons and holes. Anomaly at around ~100 K, however, always exist with gate or zero gate and also with field or no field, suggesting a robust feature in this system.

The observed resistivity behavior is consistent with temperature-dependent transport parameters such as carrier density $n$ and Hall mobility $\mu$, as shown in **Figure 3**(c). The transport parameters were extracted from longitudinal and transverse resistance data recorded during magnetic field-sweeps at different temperatures. The zero-field resistivity obtained from field sweep data is also plotted in **Figure 3**(b). The carrier density, estimated by using the period of SdH oscillations,[62] is also shown. The carrier density falls to the lowest value of $n \sim 1.15 \times 10^{11}$ cm$^{-2}$ at around 110 K where the zero-field resistivity peaks to its maximum. Note that the zero-field resistivity peak is shifted by ~20 K lower as compared to the resistivity measured by continuously ramping the temperature, which may be caused by actual temperature variation on the sample. The two temperature dependent resistivity measurements at 0 ant 5 T confirm that changing temperature can also tune graphene Fermi level across the Dirac point, similar to the electric field tuning of the graphene Fermi-level. The Hall mobility increases from $\mu \sim 1.1 \times 10^4$ cm$^2$/V·s at 2 K ($n_{2K} \sim 1.8 \times 10^{12}$ cm$^{-2}$) to a maximum value of $\mu \sim 2.6 \times 10^4$ cm$^2$/V·s at 100 K ($n_{100\,K} \sim 1 \times 10^{11}$ cm$^{-2}$) and again decreases to a value of $\mu \sim 1.2 \times 10^4$ cm$^2$/V·s at RT ($n_{300\,K} \sim 7 \times 10^{11}$ cm$^{-2}$). It is interesting to note that, taking into account the shift in *n* with temperature, the mobility away from the Dirac point is almost completely independent of temperature up to RT. The carrier density dependence of mobility over an entire temperature range (2 – 300 K), however, scales approximately as $n \propto 1/\mu$ (inset of **Figure 3**(b)), which is very similar to the gate dependence mobility observed in the previous section rather than temperature-dependent behavior usually observed in typical graphene devices.[55, 63] Therefore, this dependence implies that charged impurity scattering still dominates in our devices up to high temperatures. As described in previous section, due to screening of short-range potential scattering, mobility is enhanced at low-density regime at around 100 K as compared to low-temperatures ($T \lesssim 50$ K) and high temperatures ($T \gtrsim 200$ K) where the density is comparatively high.

We argue that the unusual temperature dependence of the transport behavior in our device is caused by the unique interaction between graphene and the LaAlO$_3$/SrTiO$_3$; residual traps cannot induce a nearly one order of magnitude change in carrier density in graphene, otherwise the sample would have been more $p$-doped than $n$-doped at RT. The origin of temperature dependent carrier tunability across Dirac point observed in our devices is attributed to the modulation doping in graphene due to induced change of carrier density at the LaAlO$_3$/SrTiO$_3$ interface as a function of temperature. Early studies showed that carrier density of 2-dimensional electron gas (2DEG) formed at the LaAlO$_3$/SrTiO$_3$ interface changes by $\sim 1 \times 10^{12}$ cm$^{-2}$ when a semi-insulating, 3 unit cells (uc) LaAlO$_3$/SrTiO$_3$ sample is cooled from 300 K to 20 K.[64] The 2DEG density increases sharply up to 2-orders of magnitude when the LaAlO$_3$ coverage is increased only half a unit cell to 3.5 uc.[32] Moreover, it should be pointed out that the maximum change of carrier density in graphene is about $\sim 1 \times 10^{12}$ cm$^{-2}$, which is almost the same order of magnitude as in the case for oxide interface. It should also be noted that the 2DEG carrier density at the LaAlO$_3$/SrTiO$_3$ interface is very sensitive to any modifications on the surface and lattice structure, and, therefore, transport properties of graphene may change correspondingly. Signatures of those modifications are also reflected by our results, in which graphene resistance shows anomaly at around $\sim 100$ K due to the ferroelastic phase transition of SrTiO$_3$ between the tetragonal and cubic phase.[65] Such anomaly of resistance behavior has been also reported in previous work on graphene/SrTiO$_3$.[50] Therefore, substrate-induced anomaly and unique coupling with graphene qualitatively supports observed unusual temperature dependent transport behavior in our samples. Deeper understanding requires further systematic investigations, which is outside the focus of our work.

Now we investigate the carrier scattering behavior in our graphene devices via magnetotransport measurements. First we show the density dependence of magnetotransport properties at 2 K. **Figure 4**(a) shows a series of magnetoresistance (MR) data of the graphene device measured at 2 K over a broad range of carrier densities near the Dirac point. The MR is calculated by using the expression: $\Delta\rho_{xx}(B) = \rho_{xx}(B) - \rho_{xx}(B=0)$. Carrier density is estimated by simultaneously recorded Hall coefficients. Overall, our observation reveals a systematic variation of MR behavior depending on the carrier density. The MR behavior can be divided into two regimes, i.e., strong-field ($\omega_c \cdot \tau_p \gg 1$) and weak-field ($\omega_c \cdot \tau_p \lesssim 1$), where $\omega_c = 5.51 \times 10^{13} \sqrt{B}$ s$^{-1}$ is the cyclotron frequency in graphene.[66] Since momentum scattering time $\tau_p$ in our devices ranges from $0.06 - 0.1$ ps for high to low carrier density regimes measured, we can roughly estimate the transport field limit; $\omega_c \cdot \tau_p \sim 1$ to be $B_c \sim 50 - 100$ mT. At strong-field limit ($B \gg B_c$), pronounced SdH oscillations and corresponding Hall plateaus at each oscillation minima are observed. A clear change of period (or frequency) is visible with changing gate voltages $V_{sg}$ (i.e., carrier density). An analysis of the minima of the SdH oscillations for a carrier density of $n = -1.6 \times 10^{12}$ cm$^{-2}$ yields the expected Berry phase of $\pi$ for single-layer graphene[6, 7] (Supporting Information). At weak-field limit ($B \lesssim B_c$), the MR is almost unchanged except for some random fluctuations, which might be attributed to universal conductance fluctuations[67] Such a flat MR behavior around zero-field is an indication of suppressed WL. As the carrier density is further reduced by tuning the Fermi level towards the Dirac point, the MR gradually increases near $B = 0$ T, showing a cusp-like dip. Observation of positive magnetoresistance is a clear signature of WAL in our samples. The positive MR increases faster as the carrier density approaches $n_{min}$, where the mobility is the highest, showing noticeable correlation with density and the mobility.

To further explore the scattering mechanism, the results were fitted with a theoretical model of quantum interference corrections to magnetoconductivity for graphene,[17] where three dominant scattering processes have been considered: usual inelastic (phase breaking) scattering, elastic inter-valley scattering, and intra-valley scattering. According to this model, the correction to the magnetoconductance is expressed as:

$$\Delta\sigma(B) = \frac{e^2}{\pi h}\left[F\left(\frac{B}{B_\phi}\right) - F\left(\frac{B}{B_\phi + 2B_i}\right) - 2F\left(\frac{B}{B_\phi + B_i + B_*}\right)\right], \quad (1)$$

where $F(Z) = ln(Z) + \Psi\left(\frac{1}{2} + \frac{1}{Z}\right)$, where $\Psi$ is the digamma function and $B_{\phi,i,*} = \frac{\hbar}{4eD}L_{\phi,i,*}^{-2}$ denotes characteristic magnetic fields associated with the diffusion constant $D$. Here, $L_\phi$ is the phase coherence length, $L_i$ is the elastic inter-valley scattering length, and $L_*$ is related to the intra-valley scattering length $L_{iv}$ and trigonal warping length $L_w$ via $L_*^{-2} = L_{iv}^{-2} + L_w^{-2}$. The theory assumes that momentum scattering rate $\tau_p^{-1}$ is the highest in the system and does not affect the electron interference. From the scattering lengths $L_{\phi,i,*}$ we can extract corresponding scattering rates $\tau_{\phi,i,*}^{-1}$ using the relation $L_{\phi,i,*} = \sqrt{D\tau_{\phi,i,*}}$. The first term in Equation (1) corresponds to the usual WL observed in typical 2D systems where the electron mean free path is shorter than the phase coherence length, while the second and third terms lead to WAL. The shape of the magnetoconductance (MC) curves varies because it results from the interplay between all scattering processes involved.

**Figure 4**(b) shows some of the low-field MR plots measured in the vicinity of the Dirac point. Fits to individual data by using Equation (1) are overlaid onto the experimental curves as thin solid lines. To avoid possible ambiguities in fitting the MC data, we assumed inter-valley scattering to be much smaller than other scattering terms, and we considered it to be constant, i.e., $B_i = 0.0001$ ($\tau_i \sim 2.5 - 5$ ps) for each density dependent data, which is plausible for strong WAL signal measured in our experiment. Fitting was performed by making sure that the condition $\tau_* < \tau_\phi$ is satisfied. We have also, for comparison sake, followed other fitting approaches as discussed in previous works,[23, 68] and obtained qualitatively similar results. However, fitting with constant $B_i$ shows better fitting results up to well-above

the transport field-limit, i.e., $B \sim \pm 150$ mT $> B_c$. We have determined two free parameters $B_{\phi,*}$ from the fit and corresponding scattering parameters have been extracted, which are plotted in **Figure 4**(b) and (c). Scattering lengths $L_{\phi,*}$ are extracted directly by using the expression $B_{\phi,*} = \frac{\hbar}{4eD} L_{\phi,*}^{-2}$. To extract the corresponding scattering rates $\tau_{\phi,*}^{-1}$, we have estimated the diffusion constant by using the relation $D = (v_F \, l_{mfp})/2 = (v_F/2) \cdot (h/(2e^2 k_F R_{xx}))$, where $v_F$ is the Fermi velocity, $k_F$ is the Fermi wave vector, and $l_{mfp}$ is the carrier mean free path. The Fermi wave vector $k_F$ in graphene is related to the carrier density $n$ by $k_F = \sqrt{\pi n}$. We determine $k_F$ from the carrier density measured by Hall measurements at 6 positions (i.e., 6 gate voltages, $V_{sg}$) in the vicinity of the Dirac point. Then we calculate diffusion constants using the typical Fermi velocity in graphene, $v_F = 1 \times 10^8$ cm/s, which also allow us to convert scattering rates using the expression described earlier. **Figure 4**(c) shows the extracted scattering lengths $L_{\phi,*}$ and rates $\tau_{\phi,*}^{-1}$ plotted as a function of carrier density. The dephasing length $L_\phi$ and the rate $\tau_\phi^{-1}$ both show minor changes above the density of $n_t > 6 \times 10^{10}$ cm$^{-2}$ but rapidly changes, i.e., the length increases and the rate drops, at the density $n_t < 6 \times 10^{10}$ cm$^{-2}$. The length increases from ~200 nm to ~400 nm and the rate drops from ~7 ps$^{-1}$ to ~1 ps$^{-1}$ over the entire range measured. The elastic scattering length $L_*$ and rate $\tau_*^{-1}$, however, weakly follow the similar trend. The observation of a rapid changes of scattering parameters near the Dirac point can be understood from the density dependence of mobility of our samples, as shown in the inset of **Figure 2**(d). Since the scattering associated with short-range defects is stronger at high carrier density, one would expect WL due to dominant inter-valley scattering process. Due to strong screening of short-range charged impurity scattering close to the Dirac point, chirality breaking inter-valley scattering is suppressed and dominated by long-range intra-valley process, showing noticeable transition towards strong WAL. This observation qualitatively agrees also with predicted favorable WAL regime in graphene[17] because ratio of scattering times $\tau_\phi/\tau_*$ gets smaller at around $n < n_t$ due to decrease of dephasing rate by reducing the carrier density. In contrast to previous observations,[19, 69] in which the dephasing length decreases as the carrier density decreases, our results suggest low scattering contribution from residual electrons and holes around the Dirac point and significant screening of inter-valley scattering processes, further highlighting the essential role of the LaAlO$_3$/SrTiO$_3$ substrate.

**Figure 5**(a) shows MC, $\Delta\sigma_{xx}$ measured while sweeping the magnetic field from $B = -5$ T to $B = +5$ T at different temperatures. The MC data were calculated from the measured MR data using the expression: $\Delta\sigma_{xx}(B) = \rho_{xx}^{-1}(B) - \rho_{xx}^{-1}(B = 0)$. For better visibility, each curve except 40 K curve is vertically shifted. MC data below 40 K is excluded because they show almost flat MC around zero field and pronounced SdH oscillations at higher fields (Supporting Information). At 50 K, begin to see a broad peak near $B = 0$ T and subsequent negative MC. The width and the relative change of the peak increase with temperature and reach a maximum at $\sim 100 - 120$ K where the carrier density reaches the lowest value (the mobility reaches the highest). Above 120 K, the dip decreases and widens again as the carrier density slowly increases and the mobility slowly decreases. Interestingly, it turns out that the negative MC deep persists up to 300 K, providing strong evidence that the WAL quantum interference survives even at RT.

To examine the WAL further, weak-field MC data is fitted using a theoretical model described by Equation (1) and scattering lengths and rates are extracted. The fitting was performed applying the same approach as described in previous section, in which we assumed inter-valley scattering term to be much smaller than dephasing and intra-valley terms and unchanged, i.e., $B_i = 0.001$ ($\tau_i \sim 6$ ps) for all temperature dependent data. This is equally plausible as in the density dependent data because of strong WAL signal measured in our experiment. Plots of WAL fits are shown in **Figure 5**(b) with four representative data (symbols) with fits (solid lines with respective symbol colors). The model fits reasonably well for all temperatures. **Figure 5**(c) shows the extracted scattering parameters plotted as a function of temperatures. The overall trend shows that the dephasing length $L_\phi$ decreases and the dephasing rate $\tau_\phi^{-1}$ increases with increasing the temperature. However, some anomaly occurs in the vicinity of Dirac point at around 110 K, where dephasing length $L_\phi$ sharply increases and the rate $\tau_\phi^{-1}$ sharply drops. The maximum dephasing length of $L_\phi \sim 200$ nm and the minimum dephasing rate of $\tau_\phi^{-1} \sim 1$ ps$^{-1}$ are achieved near 100 K; and both changes to $L_\phi \sim 60$ nm and $\tau_\phi^{-1} = 16$ ps$^{-1}$, respectively, at RT. We can attribute to the anomaly with a similar argument as discussed in the previous section, which is caused by direct correlation with the change to carrier density and the mobility around the Dirac point and supports our explanation of greatly suppressed short-range scattering due to enhanced dielectric screening in our sample. The intra-valley scattering length $L_*$ and the rate $\tau_*^{-1}$, however, remain unchanged, as expected, except minor changes at around 110 K. When we consider the data only outside the Dirac point, i.e., $T \geq 130$ K, dephasing rate $\tau_\phi^{-1}$ roughly proportional to temperature and shows much higher values than predicted transport-phonon scattering rate.[24] Earlier studies[70, 71] [24, 72] have found electron-electron scattering will follow linear temperature dependence of dephasing rate $\tau_\phi^{-1}$ at low temperatures in diffusive regime, i.e., $k_B T \tau_p / \hbar < 1$ and parabolic dependence at high temperatures in ballistic regime, $k_B T \tau_p / \hbar > 1$. In our samples the parameter $k_B T \tau_p / \hbar$ varies from $\sim 1 - 5$ in the temperature range of $\sim 80 - 300$ K, which confirms dominantly a ballistic regime. Our results show, in deed, the dephasing rate qualitatively obey the usual parabolic temperature dependence for ballistic regime, as described by the expression,[24, 73]

$$\tau_\phi^{-1} \approx \beta \cdot \frac{\pi}{4} \cdot \left\{ \frac{(k_B T)^2}{\hbar E_f} \right\} \cdot \ln\left(\frac{2E_f}{k_B T}\right), \qquad (2)$$

with $\beta \sim 1$, the empirical coefficient, attributing to the dominant contribution of electron-electron scattering in the decoherence process. Since $L_\phi = \sqrt{D/\tau_\phi^{-1}}$, dephasing length $L_\phi$ also follows the inverse relationship with temperature, i.e., $L_\phi \propto 1/T$. Except a deviation around anomaly observed in between $100 - 130$ K, model agrees surprisingly well even with taking account the change of carrier density with temperature, as shown by solid black lines in **Figure 5**(c). Earlier localization studies on graphene have also found similar results at relatively lower temperatures,[18, 19, 23, 24] which are explained by diffusive electron-electron interaction.[70, 71] Therefore, it is interesting to see that the electron-electron scattering is still a dominating dephasing source in our graphene devices up to RT, which we assume as a result of suppressed phonon contribution due to unique coupling with the substrate.

WAL is predicted in pristine and defect-free single layer graphene[74, 75] and is attributed to the signature of pseudospin, arising from the symmetry of the honeycomb graphene lattice. This can be understood in terms of nontrivial Berry phase of $\pi$ induced by pseudospin rotation, which is aligned with momentum in a single layer graphene. This causes a phase change of $\pi$ at each backscattering process and, as a consequence, a suppression of backscattering is expected. In a bilayer, however, the electronic structure changes dramatically in which the low-energy quasiparticle band dispersion becomes quadratic as in a conventional 2DEG system and with $2\pi$-Berry phase the pseudospin turns twice as quickly in the plane than momentum.[8] As a result, no suppression of backscattering is expected and this has been experimentally verified.[76] Despite the theoretical prediction, evidence of WAL quantum interference signature has been reported only in exfoliated and epitaxial graphene samples.[23, 24] WL has been routinely observed in different kinds of graphene samples,[19, 77] and is attributed to presence of disorder. In contrast to previous studies of CVD-grown graphene on SiO$_2$, our results demonstrate the first direct evidence of WAL observed in CVD-grown graphene samples and point to the critical role of the ultraflat surface and conducting interface of the LaAlO$_3$/SrTiO$_3$ substrate in suppressing short-range inter-valley scattering. In addition, we have observed a crossover from WL to WAL regimes near the Dirac point at 2 K as the carrier density is reduced and mobility is increased by tuning the Fermi level, a consequence of enhanced dielectric screening of short-range charged impurity scattering close to the Dirac point thereby suppressing the inter-valley scattering contributions. As we have shown in **Figure 4**(a), WAL is completely suppressed in the high-density regime, where graphene's own screening ability is activated due to abundant carriers and consequently inter-valley scattering dominates. Therefore, the interplay of inter-valley and intra-valley scattering contributions and the role of substrate dielectric screening qualitatively supports our observation of a systematic transition from WL to WAL as the Fermi level is tuned towards the Dirac point.

Temperature dependent results are quite interesting in our sample because of unusual temperature dependent transport behavior. As shown in **Figure 3**, our devices show ambipolar tuning of carrier density and mobility with temperature, which is correlated with the field-effect gating behavior. Observation of SdH oscillations, QHE, and clear flip of carriers from holes to electrons at higher temperatures reflects unique coupling with the substrate, which we attribute to a modulation doping of graphene through the substrate interface. Most important result of our work is the observation of WAL quantum interference up to RT, which we explain based on our quantitative analysis of WAL data. Due to increased dephasing rate $\tau_\phi^{-1}$ and relatively constant intra-valley elastic scattering rate $\tau_*^{-1}$, the ratio of $\tau_\phi/\tau_*$ is always smaller ($\leq 2.5$) than or comparable with the ratio at 2 K, which meets favorable transport conditions for the observation of WAL as predicted by theory.[17] Moreover, electron-phonon scattering in graphene is predicted to be very weak[78, 79] as compared to typical metallic and semiconducting 2D systems and, therefore, it is expected that quantum interference may exist in graphene up to high temperatures. Earlier transport studies on exfoliated graphene samples on $SiO_2$ observed WAL effect up to 200 K and beyond that it disappears due to rapid dephasing of quantum interference between time-reversed electron-trajectories.[24] Therefore, our results suggest critical role of suppressed phonon contribution in our sample up to RT due to unique interaction with the substrate. Strong evidence of this effect can be observed in **Figure 5**(c), which shows experimentally extracted dephasing rate agrees well with electron-electron scattering rate obtained from theory up to RT, attributing to dominant contribution of electron-electron scattering and suppressed phonon contribution in the decoherence process. Good agreement of inelastic scattering mechanism with electron-electron scattering in a purely ballistic regime is a rare test, which has been captured in our experiment. In addition, the RT mobility of our devices is $\mu(300\ \text{K}) \sim 1.2\times 10^4$ cm$^2$/V·s, which is as good as the mobility at low-temperatures for similar carrier density, again suggesting weak phonon contribution. Another point to add is low-doping concentration in our samples at RT due to good interface quality, which has positive consequences for substrate-induced screening of short-range charged impurity scattering. Therefore, strong dielectric screening of charged impurity scattering, minimal surface corrugation, good interface quality, and suppression of phonon scattering all have played a crucial role for the long dephasing length over a broad temperature range in our samples, which allows the preservation of quantum interference phenomena of WAL up to RT.

Alternative mechanisms for WAL signatures have also been considered. Proximity-induced spin-orbit coupling (SOC) in graphene through the interface[80, 81] can lead to a pronounced WAL.[81] Previous studies reported[39, 40] a gate-tunable Rashba-like SOC at the conducting interface between LaAlO$_3$ and SrTiO$_3$. This effect relates to conductivity of a distinct electronic system that is well isolated by the wide bandgap of LaAlO$_3$ layer from the graphene layer. Besides that, electron-electron interaction can also lead to negative MC in a moderate field regime in graphene, as reported in previous works.[22, 82-84] Since our work focused mainly in a weak-field regime and our results agree well with existing theory of quantum interference correction in graphene, we believe, WAL observed in our samples is, indeed, a consequence of pseudospin quantum interference phenomena.

Very similar results of positive MR around zero-field have been reported in a high mobility metallic 2D system and the results were explained considering classical partial-diffuse scattering effect at the device boundary.[85] However, we need to consider that the metallic system discussed in this paper is fundamentally different from graphene. The unique pseudospin texture of Dirac electrons leads to various quantum phenomena such as half-integer quantum Hall effect (QHE) in graphene. It has been demonstrated by earlier work[86] that quantum phenomena can exist in graphene even at RT (at very high magnetic fields). We have also observed QHE in our samples at the temperature as high as ∼ 100 K (see also Supporting Information) using relatively low magnetic fields. Therefore, based on our observation we conclude that it is indeed quantum interference phenomena what we observed up to RT.

In conclusion, we have achieved high mobility graphene field-effect devices by integrating CVD-grown graphene with the complex-oxide heterostructure $LaAlO_3/SrTiO_3$. These devices show anomalous quantum Hall effect and suppressed backscattering of quantum interference signature, as expected from high quality graphene samples. Unique features of this system are (1) the sharp rise of carrier mobility close to the Dirac point attributed to screening of short-range charged impurity scattering by the $LaAlO_3/SrTiO_3$ interface and (2) temperature-independent high mobility up to 300 K due to weak phonon scattering contribution. Therefore, assisted by unique interaction with oxide substrate, this work represents first direct observation of suppressed-backscattering as WAL quantum interference signature in CVD-grown graphene at both low temperatures (∼2 K) and high temperatures (∼300 K). The persistence of WAL to RT is an important step forward to the realization of pseudospin-based graphene devices, and presents an opportunity to explore novel physics that might result from proximal coupling with oxide interface. This will also open up new opportunities for the fabrication of multifunctional devices that combine graphene with a complex-oxide herointerface.

# Experimental Section

**LaAlO$_3$/SrTiO$_3$ sample preparation:** The LaAlO$_3$/SrTiO$_3$ heterostructure used in this study was fabricated by growing 3.4 unit cell (~1.3 nm) LaAlO$_3$ film on TiO$_2$-terminated SrTiO$_3$ (001) substrate by using pulsed laser deposition (PLD).[87, 88] The LaAlO$_3$ thickness was precisely controlled by in-situ monitoring of the reflection high-energy electron diffraction (RHEED) intensity during the layer-by-layer growth of LaAlO$_3$. Active device regions were created on a LaAlO$_3$/SrTiO$_3$ sample by fabricating gold (Au) electrodes lines and contact pads. The electrodes were made using a two-step deposition process: first, interface-connected contacts were made via Ar+ etching followed by Ti/Au (4 nm/25 nm) sputter deposition and second, direct patterning of Au electrodes on the surface of LaAlO$_3$/SrTiO$_3$. Second electrodes connect to the interface connected Au electrodes at one end and act as contact pads for wire bonding at the other end. In our sample we have fabricated 8 top electrodes and 8 electrodes contacting the interface. The samples were mounted on a chip carrier and electrical connections were made by wire bonding between chip carrier and Au contact pads. A background resistance between two Au electrodes is $> 1$ G$\Omega$ at 300 K, which confirms the expected insulating ground state of the 3.4 unit cells LaAlO$_3$/SrTiO$_3$ samples.

**Growth, transfer, and patterning of graphene on LaAlO$_3$/SrTiO$_3$:** Large-area, single-layer graphene samples examined in this study were synthesized by atmospheric pressure chemical vapor deposition (APCVD) method on Cu substrates.[44] Following the synthesis, graphene layer of approximately 10 mm × 10 mm size were transferred assisted by poly(methyl methacrylate) (PMMA) onto pre-patterned LaAlO$_3$/SrTiO$_3$ substrates using a wet-transfer procedure in which the Cu substrate was removed by etching in ~1M ammonium persulphate solution. Prior to the graphene transfer process, the surface of the LaAlO$_3$/SrTiO$_3$ substrate was cleaned in an oxygen plasma to minimize possible contamination between graphene and LaAlO$_3$ surface. Deep-UV lithography and oxygen plasma cleaning were used to selectively remove unwanted graphene. After removal, the remaining circular graphene piece was aligned with the LaAlO$_3$/SrTiO$_3$ canvases with gold electrodes that independently contact the graphene on the surface and LaAlO$_3$/SrTiO$_3$ at the interface. Using c-AFM lithography, graphene samples were patterned into Hall bars with a nominal channel length of $L = 3$ μm and width of $W = 5$ μm, as measured from the center of the voltage probes,.

**Sample characterization:** An AFM (Asylum MFP-3D) was used to characterize the surface morphology of the transferred graphene. AFM images were acquired in air using silicon cantilevers operated in tapping mode. Surface roughness was characterized as the standard deviation of the surface height distribution (see Supporting Information). Raman spectroscopy was used to characterize doping and defects in graphene. The measurements were performed using Raman microscope setup (Renishaw InVia) with 633-nm laser excitation under ambient conditions. Details of Raman spectroscopy results are described in the Supporting Information. Transport measurements were performed in a four-terminal geometry using standard lock-in techniques at $\sim 1-10$ Hz. Samples were cooled in a variable-temperature $(2-300$ K$)$ liquid $^4$He flow cryostat.

## Supporting Information

Supporting information is available online from the Wiley Online Library or from the author.

## Acknowledgements

This work is supported by ONR GCO N00014-13-1-0806 (C.B.E. and J.L.), AFOSR FA9550-12-1-0342 (C.B.E.) and FA9550-15-1-0334 (C.B.E.), and the National Science Foundation DMR- 1104191 (J.L.).

# References

[1] E. Abrahams, P. W. Anderson, D. C. Licciardello, T. V. Ramakrishnan, Phys. Rev. Lett. 1979, 42, 673.

[2] S. Hikami, A. I. Larkin, Y. Nagaoka, Progr. Theor. Phys. 1980, 63, 707.

[3] G. Bergman, Phys. Rev. Lett. 1982, 48, 1046.

[4] G. Bergmann, Phys. Rep. 1984, 107, 1.

[5] S. Konschuh, M. Gmitra, J. Fabian, Phys. Rev. B 2010, 82, 245412.

[6] Y. B. Zhang, Y. W. Tan, H. L. Stormer, P. Kim, Nature 2005, 438, 201.

[7] K. S. Novoselov, A. K. Geim, S. V. Morozov, D. Jiang, M. I. Katsnelson, I. V. Grigorieva, S. V. Dubonos, A. A. Firsov, Nature 2005, 438, 197.

[8] K. S. Novoselov, E. McCann, S. V. Morozov, V. I. Fal'ko, M. I. Katsnelson, U. Zeitler, D. Jiang, F. Schedin, A. K. Geim, Nat. Phys. 2006, 2, 177.

[9] M. I. Katsnelson, K. S. Novoselov, A. K. Geim, Nat. Phys. 2006, 2, 620.

[10] A. F. Young, P. Kim, Nat. Phys. 2009, 5, 222.

[11] N. Stander, B. Huard, D. Goldhaber-Gordon, Phys. Rev. Lett. 2009, 102, 026807.

[12] A. Rycerz, J. Tworzydlo, C. W. J. Beenakker, Nat. Phys. 2007, 3, 172.

[13] P. San-Jose, E. Prada, E. McCann, H. Schomerus, Phys. Rev. Lett. 2009, 102, 247204.

[14] D. Pesin, A. H. MacDonald, Nat. Mater. 2012, 11, 409.

[15] S. K. Banerjee, L. F. Register, E. Tutuc, D. Reddy, A. H. MacDonald, IEEE Electron Device Lett. 2009, 30, 158.

[16] S. Roche, J. Aring;kerman, B. Beschoten, J. C. Charlier, M. Chshiev, S. P. Dash, B. Dlubak, J. Fabian, A. Fert, M. Guimaraes, F. Guinea, I. Grigorieva, C. Schonenberger, P. Seneor, C. Stampfer, S. O. Valenzuela, X. Waintal, B. van Wees, 2D Mater. 2015, 2, 030202.

[17] E. McCann, K. Kechedzhi, V. I. Fal'ko, H. Suzuura, T. Ando, B. L. Altshuler, Phys. Rev. Lett. 2006, 97, 146805.

[18] S. V. Morozov, K. S. Novoselov, M. I. Katsnelson, F. Schedin, L. A. Ponomarenko, D. Jiang, A. K. Geim, Phys. Rev. Lett. 2006, 97, 016801.

[19] F. V. Tikhonenko, D. W. Horsell, R. V. Gorbachev, A. K. Savchenko, Phys. Rev. Lett. 2008, 100, 056802.

[20] D. W. Horsell, F. V. Tikhonenko, R. V. Gorbachev, A. K. Savchenko, Phil. Trans. R. Soc. A 2008, 366, 245.

[21] S. Pezzini, C. Cobaleda, E. Diez, V. Bellani, Phys. Rev. B 2012, 85, 165451.

[22] B. Jouault, B. Jabakhanji, N. Camara, W. Desrat, C. Consejo, J. Camassel, Phys. Rev. B 2011, 83, 195417.

[23] X. S. Wu, X. B. Li, Z. M. Song, C. Berger, W. A. de Heer, Phys. Rev. Lett. 2007, 98, 136801.

[24] F. V. Tikhonenko, A. A. Kozikov, A. K. Savchenko, R. V. Gorbachev, Phys. Rev. Lett. 2009, 103, 226801.

[25] D. V. Tuan, F. Ortmann, D. Soriano, S. O. Valenzuela, S. Roche, Nat. Phys. 2014, 10, 857.

[26] D. V. Tuan, S. Roche, Phys. Rev. Lett. 2016, 116, 106601.

[27] T. Ando, J. Phys. Soc. Jpn. 2006, 75, 074716.

[28] E. H. Hwang, S. Adam, S. Das Sarma, Phys. Rev. Lett. 2007, 98, 186806.

[29] C. R. Dean, A. F. Young, MericI, LeeC, WangL, SorgenfreiS, WatanabeK, TaniguchiT, KimP, K. L. Shepard, HoneJ, Nat. Nanotechnol. 2010, 5, 722.

[30] Y. Zheng, G.-X. Ni, C.-T. Toh, M.-G. Zeng, S.-T. Chen, K. Yao, B. Özyilmaz, Appl. Phys. Lett. 2009, 94, 163505.

[31] A. Ohtomo, H. Y. Hwang, Nature 2004, 427, 423.

[32] S. Thiel, G. Hammerl, A. Schmehl, C. W. Schneider, J. Mannhart, Science 2006, 313, 1942.

[33] N. Reyren, S. Thiel, A. D. Caviglia, L. F. Kourkoutis, G. Hammerl, C. Richter, C. W. Schneider, T. Kopp, A. S. Ruetschi, D. Jaccard, M. Gabay, D. A. Muller, J. M. Triscone, J. Mannhart, Science 2007, 317, 1196.

[34] A. D. Caviglia, S. Gariglio, N. Reyren, D. Jaccard, T. Schneider, M. Gabay, S. Thiel, G. Hammerl, J. Mannhart, J. M. Triscone, Nature 2008, 456, 624.

[35] A. Brinkman, M. Huijben, M. Van Zalk, J. Huijben, U. Zeitler, J. C. Maan, W. G. Van der Wiel, G. Rijnders, D. H. A. Blank, H. Hilgenkamp, Nat. Mater. 2007, 6, 493.

[36] D. A. Dikin, M. Mehta, C. W. Bark, C. M. Folkman, C. B. Eom, V. Chandrasekhar, Phys. Rev. Lett. 2011, 107, 056802.

[37] L. Li, C. Richter, J. Mannhart, R. C. Ashoori, Nat. Phys. 2011, 7, 762.

[38] F. Bi, M. C. Huang, S. Ryu, H. Lee, C. W. Bark, C. B. Eom, P. Irvin, J. Levy, Nat Commun 2014, 5, 5019.

[39] M. Ben Shalom, M. Sachs, D. Rakhmilevitch, A. Palevski, Y. Dagan, Phys. Rev. Lett. 2010, 104, 126802.

[40] A. D. Caviglia, M. Gabay, S. Gariglio, N. Reyren, C. Cancellieri, J. M. Triscone, Phys. Rev. Lett. 2010, 104, 126803.

[41] C. Cen, S. Thiel, G. Hammerl, C. W. Schneider, K. E. Andersen, C. S. Hellberg, J. Mannhart, J. Levy, Nat. Mater. 2008, 7, 298.

[42] C. Cen, S. Thiel, J. Mannhart, J. Levy, Science 2009, 323, 1026.

[43]	M. Huang, G. Jnawali, J.-F. Hsu, S. Dhingra, H. Lee, S. Ryu, F. Bi, F. Ghahari, J. Ravichandran, L. Chen, P. Kim, C.-B. Eom, B. D'Urso, P. Irvin, J. Levy, APL Mater. 2015, 3, 062502.

[44]	I. Vlassiouk, M. Regmi, P. Fulvio, S. Dai, P. Datskos, G. Eres, S. Smirnov, ACS Nano 2011, 5, 6069.

[45]	S. Dhingra, J.-F. Hsu, I. Vlassiouk, B. D'Urso, Carbon 2014, 69, 188.

[46]	J.-F. Hsu, S. Dhingra, G. Jnawali, M. Huang, F. Bi, L. Chen, P. Irvin, C.-B. Eom\, J. Levy, B. R. D'Urso, in preparation.

[47]	A. J. M. Giesbers, U. Zeitler, S. Neubeck, F. Freitag, K. S. Novoselov, J. C. Maan, Solid State Commun. 2008, 147, 366.

[48]	L. Weng, L. Zhang, Y. P. Chen, L. P. Rokhinson, Appl. Phys. Lett. 2008, 93, 093107.

[49]	X. Hong, J. Hoffman, A. Posadas, K. Zou, C. H. Ahn, J. Zhu, Appl. Phys. Lett. 2010, 97, 033114.

[50]	S. Saha, O. Kahya, M. Jaiswal, A. Srivastava, A. Annadi, J. Balakrishnan, A. Pachoud, C. T. Toh, B. H. Hong, J. H. Ahn, T. Venkatesan, B. Ozyilmaz, Sci. Rep. 2014, 4, 6173.

[51]	R. Sachs, Z. S. Lin, J. Shi, Sci. Rep. 2014, 4, 3657.

[52]	H. M. Wang, Y. H. Wu, C. X. Cong, J. Z. Shang, T. Yu, Acs Nano 2010, 4, 7221.

[53]	H. Xu, Y. Chen, J. Zhang, H. Zhang, Small 2012, 8, 2833.

[54]	J. Martin, N. Akerman, G. Ulbricht, T. Lohmann, J. H. Smet, K. Von Klitzing, A. Yacoby, Nat. Phys. 2008, 4, 144.

[55]	W. J. Zhu, V. Perebeinos, M. Freitag, P. Avouris, Phys. Rev. B 2009, 80, 235402.

[56]	J. H. Chen, C. Jang, S. Adam, M. S. Fuhrer, E. D. Williams, M. Ishigami, Nat. Phys. 2008, 4, 377.

[57]	X. Hong, A. Posadas, K. Zou, C. H. Ahn, J. Zhu, Phys. Rev. Lett. 2009, 102, 136808.

[58]	M. H. Yusuf, B. Nielsen, M. Dawber, X. Du, Nano Lett. 2014, 14, 5437.

[59]	X. Li, C. W. Magnuson, A. Venugopal, J. An, J. W. Suk, B. Han, M. Borysiak, W. Cai, A. Velamakanni, Y. Zhu, L. Fu, E. M. Vogel, E. Voelkl, L. Colombo, R. S. Ruoff, Nano Lett. 2010, 10, 4328.

[60]	W. Gannett, W. Regan, K. Watanabe, T. Taniguchi, M. F. Crommie, A. Zettl, Appl. Phys. Lett. 2011, 98, 242105.

[61]	E. H. Hwang, S. Adam, S. D. Sarma, Phys. Rev. Lett. 2007, 98, 186806.

[62]	Y. Z. Andrea F. Young, and Philip Kim, *Physics of Graphene*, Springer International Publishing, 2014.

[63]	J. H. Chen, C. Jang, S. D. Xiao, M. Ishigami, M. S. Fuhrer, Nat. Nanotechnol. 2008, 3, 206.

[64]	J. S. Lee, S. K. Seung, S. R. Lee, J.-W. Chang, H. Noh, L. Baasandorj, H. S. Shin, S.-B. Shim, J. Song, J. Kim, Phys. Status Solidi (RRL) 2012, 6, 472.

[65] K. A. Müller, T. W. Kool, *Properties of Perovskites and Other Oxides*, World Scientific Publishing Co. Pte. Ltd., 2010.

[66] S. Das Sarma, S. Adam, E. H. Hwang, E. Rossi, Rev. Mod. Phys. 2011, 83, 407.

[67] C. Berger, Z. Song, X. Li, X. Wu, N. Brown, C. Naud, xe, cile, D. Mayou, T. Li, J. Hass, A. N. Marchenkov, E. H. Conrad, P. N. First, W. A. de Heer, Science 2006, 312, 1191.

[68] F. Ortmann, A. Cresti, G. Montambaux, S. Roche, Europhys. Lett. 2011, 94, 47006.

[69] D.-K. Ki, D. Jeong, J.-H. Choi, H.-J. Lee, K.-S. Park, Phys. Rev. B 2008, 78, 125409.

[70] B. L. Altshuler, A. G. Aronov, D. E. Khmelnitsky, J. Phys. C: Solid State Phys. 1982, 15, 7367.

[71] E. Abrahams, P. W. Anderson, P. A. Lee, T. V. Ramakrishnan, Phys. Rev. B 1981, 24, 6783.

[72] B. N. Narozhny, G. Zala, I. L. Aleiner, Phys. Rev. B 2002, 65, 180202.

[73] B. N. Narozhny, G. Zala, I. L. Aleiner, Phys. Rev. B 2002, 65, 180202(R).

[74] H. Suzuura, T. Ando, Phys. Rev. Lett. 2002, 89, 266603.

[75] T. Ando, J. Phys. Soc. Jpn. 2004, 73, 1273.

[76] R. V. Gorbachev, F. V. Tikhonenko, A. S. Mayorov, D. W. Horsell, A. K. Savchenko, Phys. Rev. Lett. 2007, 98, 176805.

[77] A. M. R. Baker, J. A. Alexander-Webber, T. Altebaeumer, T. J. B. M. Janssen, A. Tzalenchuk, S. Lara-Avila, S. Kubatkin, R. Yakimova, C. T. Lin, L. J. Li, R. J. Nicholas, Phys. Rev. B 2012, 86, 235441.

[78] T. Stauber, N. M. R. Peres, F. Guinea, Phys. Rev. B 2007, 76, 205423.

[79] E. H. Hwang, S. Das Sarma, Phys. Rev. B 2008, 77, 115449.

[80] Z. Wang, D.-K. Ki, H. Chen, H. Berger, A. H. MacDonald, A. F. Morpurgo, Nat Commun 2015, 6, 8339.

[81] A. Avsar, J. Y. Tan, T. Taychatanapat, J. Balakrishnan, G. K. W. Koon, Y. Yeo, J. Lahiri, A. Carvalho, A. S. Rodin, E. C. T. O'Farrell, G. Eda, A. H. Castro Neto, B. Özyilmaz, Nat Commun 2014, 5, 4875.

[82] B. Jabakhanji, D. Kazazis, W. Desrat, A. Michon, M. Portail, B. Jouault, Phys. Rev. B 2014, 90, 035423.

[83] J. Jobst, D. Waldmann, I. V. Gornyi, A. D. Mirlin, H. B. Weber, Phys. Rev. Lett. 2012, 108, 106601.

[84] A. A. Kozikov, A. K. Savchenko, B. N. Narozhny, A. V. Shytov, Phys. Rev. B 2010, 82, 075424.

[85] T. J. Thornton, M. L. Roukes, A. Scherer, B. P. Van de Gaag, Phys. Rev. Lett. 1989, 63, 2128.

[86] K. S. Novoselov, Z. Jiang, Y. Zhang, S. V. Morozov, H. L. Stormer, U. Zeitler, J. C. Maan, G. S. Boebinger, P. Kim, A. K. Geim, Science 2007, 315, 1379.

[87] C. W. Bark, D. A. Felker, Y. Wang, Y. Zhang, H. W. Jang, C. M. Folkman, J. W. Park, S. H. Baek, H. Zhou, D. D. Fong, X. Q. Pan, E. Y. Tsymbal, M. S. Rzchowski, C. B. Eom, Proc. Natl. Acad. Sci. USA 2011, 108, 4720.

[88] J. W. Park, D. F. Bogorin, C. Cen, D. A. Felker, Y. Zhang, C. T. Nelson, C. W. Bark, C. M. Folkman, X. Q. Pan, M. S. Rzchowski, J. Levy, C. B. Eom, Nat Commun 2010, 1, 94.

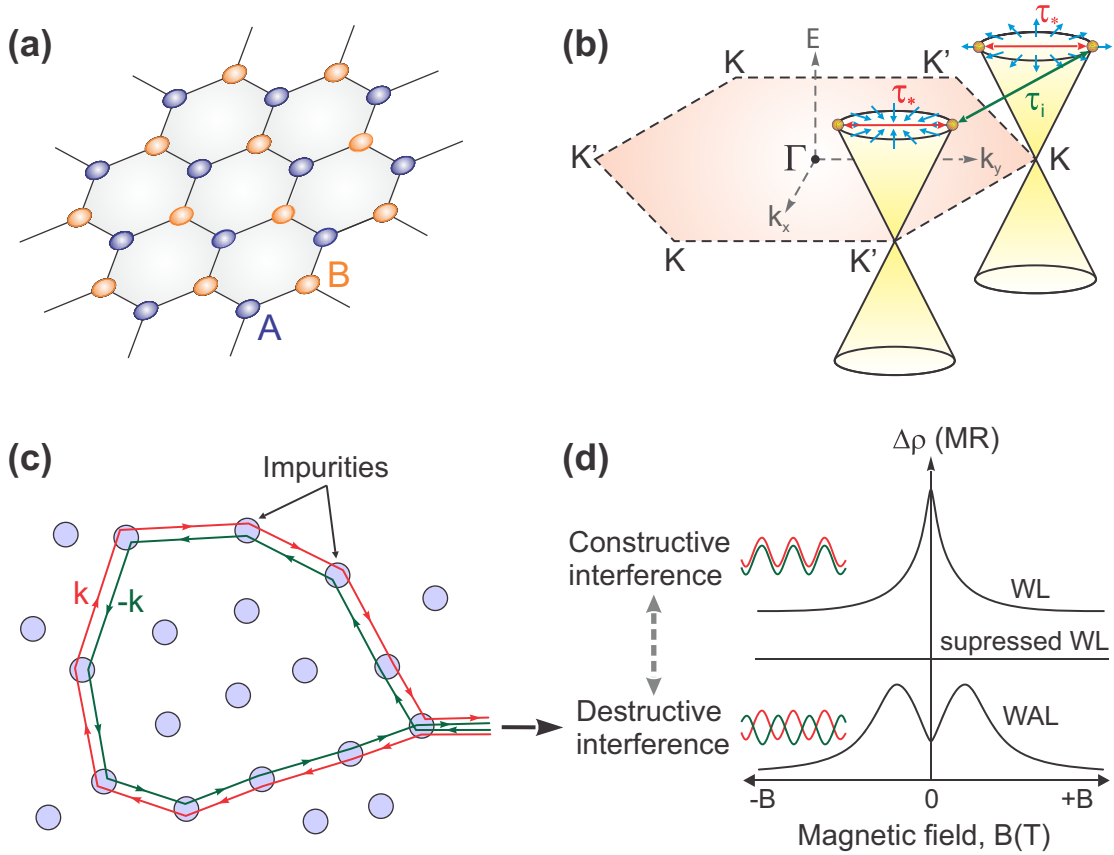

**Figure 1.** Schematic illustration of quantum interference mechanism in graphene. (a) Honeycomb lattice structure of graphene, composed of two interpenetrating triangular carbon sub-lattices denoted by "A" and "B" with two different colors. (b) The first Brillouin zone of the honeycomb lattice with its center Γ and low-energy band structure at two inequivalent corners K and K' associated with two sub-lattices. Blue arrows directing inward and outward at each Dirac cone are denoted as pseudospin degree of freedom. Solid green (red) lines indicate inter-valley (intra-valley) scattering with scattering rates, $\tau_i$ ($\tau_*$). (c) Schematic view of two time-reversed electron trajectories in a closed quantum diffusive path. (d) Typical magnetoresistance behavior in graphene due to the interplay of inter-valley and intra-valley scattering processes and the resulting interference effects. Due to the pseudospin and Berry phase of $\pi$, backscattering is suppressed via destructive interference, which leads to weak antilocalization (WAL) in high quality graphene. In the presence of short-range disorder caused by point defects,

inter-valley scattering contribution dominates, which results in suppressed weak localization or usual weak localization (WL) in moderate-quality graphene samples.

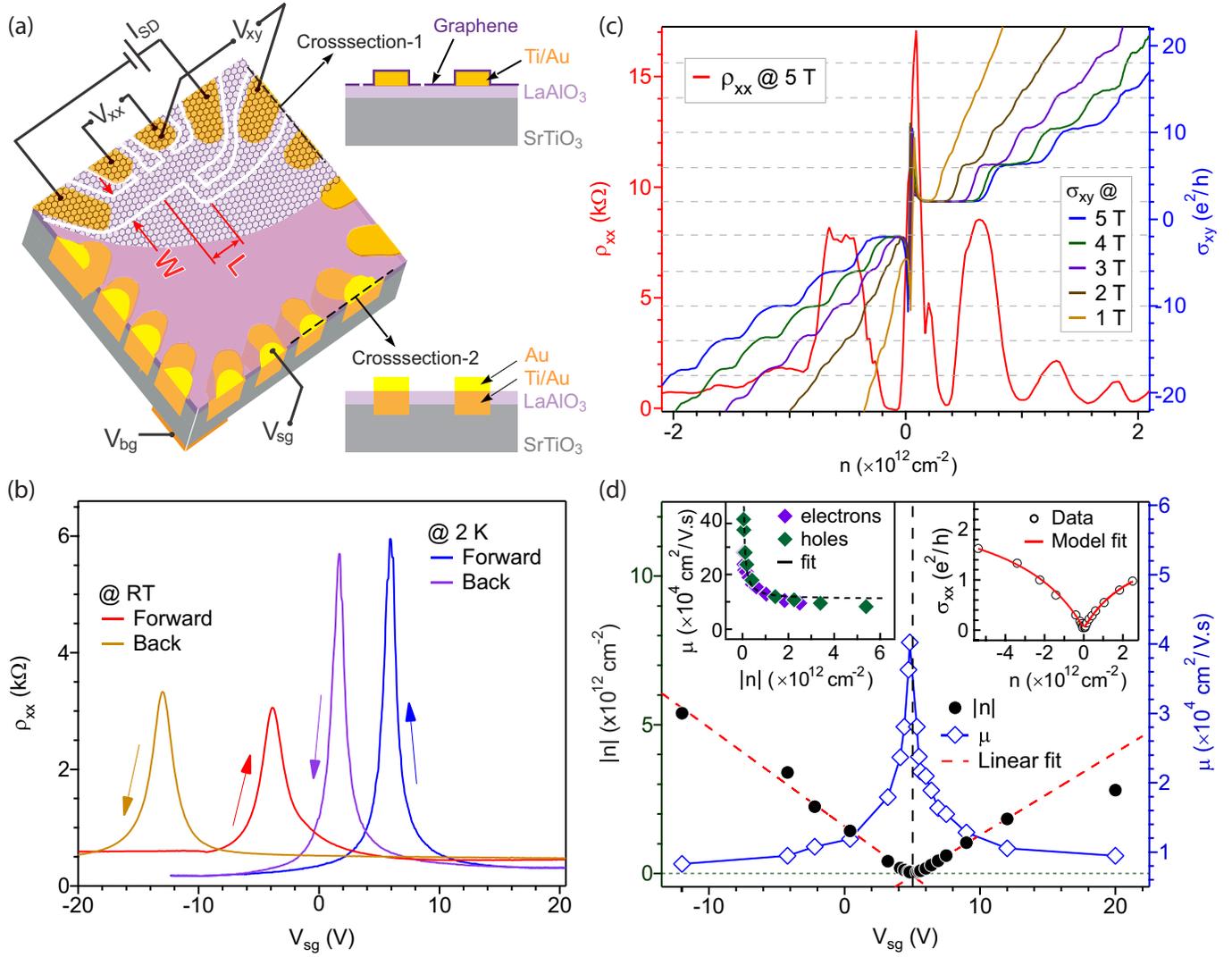

**Figure 2.** Gate-dependent transport properties. (a) Schematic view of graphene/LaAlO$_3$/SrTiO$_3$ field-effect device and transport measurement scheme. A Hall bar geometry is fabricated by selective removal of the graphene or anodic etching (indicated by white regions) using c-AFM lithography. Graphene is connected only with top-gated Au-electrodes on LaAlO$_3$/SrTiO$_3$ (upper-half), as shown in crosssection-1. Lower-half Au-electrodes are connected with interface (crossection-2), which allow gating graphene through side-gate $V_{sg}$. (b) Graphene sheet resistivity $\rho_{xx}$ as a function of applied side-gate voltages $V_{sg}$ at room temperature (RT) and 2 K. The characteristic Dirac peak is shifted as the gate sweeping direction is reversed, showing hysteretic behavior. (c) Sheet resistivity $\rho_{xx}$ and Hall conductance $\sigma_{xy}$ as a function of carrier density $n$ with perpendicular magnetic fields, showing half-integer QHE in graphene. (d) Carrier density $n$ and Hall mobility $\mu$ as a function of gate bias $V_{sg}$. Dashed red line represents a linear fit along positive and negative directions. Left inset shows the electron and hole mobilities as a function of carrier density and dashed black line is a fit with inverse relation with carrier density. Right inset shows the sheet conductivity $\sigma_{xx}$ as a function of carrier density $n$ and red dashed line is a fit to the model described in the text.

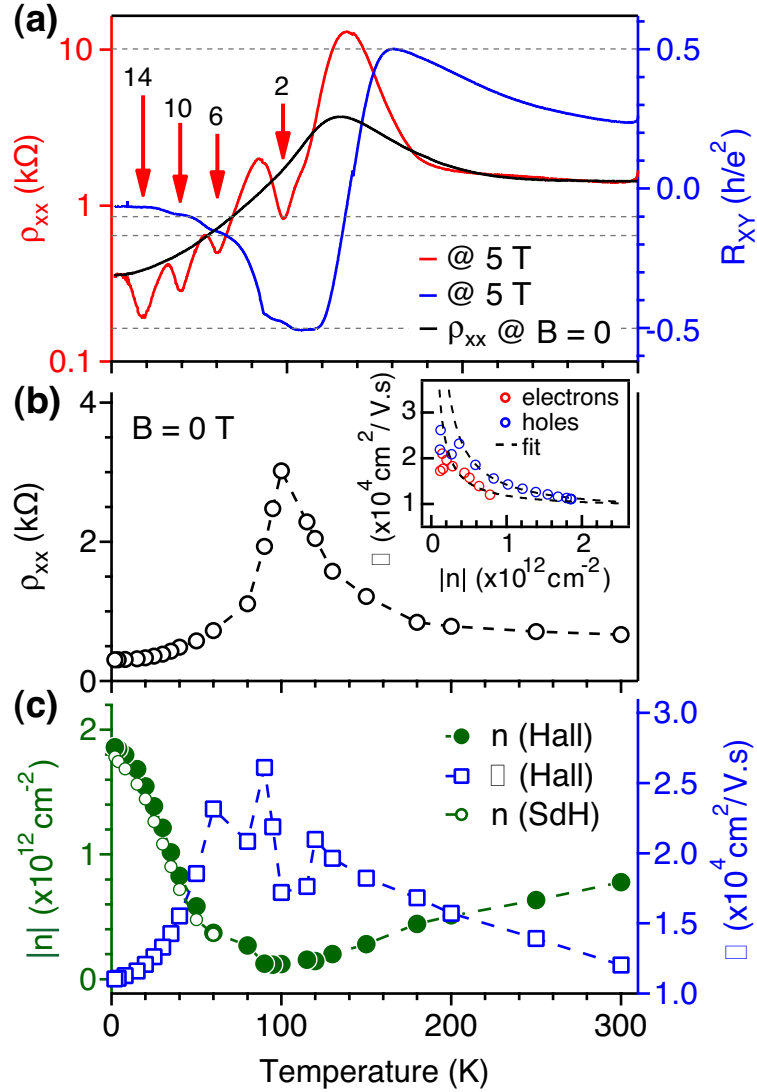

**Figure 3.** Temperature-dependent transport properties. (a) The sheet resistivity $\rho_{xx}$ and the Hall resistance $R_{xy}$ of the graphene device at $B = 5$ T while ramping the temperature from 2 K to 300 K. Zero-field sheet resistivity $\rho_{xx}(B = 0)$ is also plotted. Signature of quantum Hall plateaus, as indicated by arrows with different Landau level (LL) filling factors, $\nu = 2, 6, 10,$ and $14$ at each minimum of Shubnikov-de Haas (SdH) oscillations. Around 100 K, the polarity of the majority charge carriers changes from holes to electrons, showing temperature-dependent bipolar carrier tuning in the device. (b) Zero-field sheet resistivity $\rho_{xx}(B = 0)$ obtained from the magnetoresistance data recorded at different temperatures. The peak of $\rho_{xx}(B = 0)$ is slightly shifted compared to continuously ramping the temperatures. (c) Extracted carrier density $n$ and Hall mobility $\mu$ at different temperatures. Carrier

density $n$ was estimated from the slope of the Hall coefficient and also from the period of the SdH oscillations (up to 60 K). Inset in Figure (b) shows electron and hole mobilities plotted as a function of carrier density. Both mobilities are inversely proportional to the carrier density, as shown by black dashed lines.

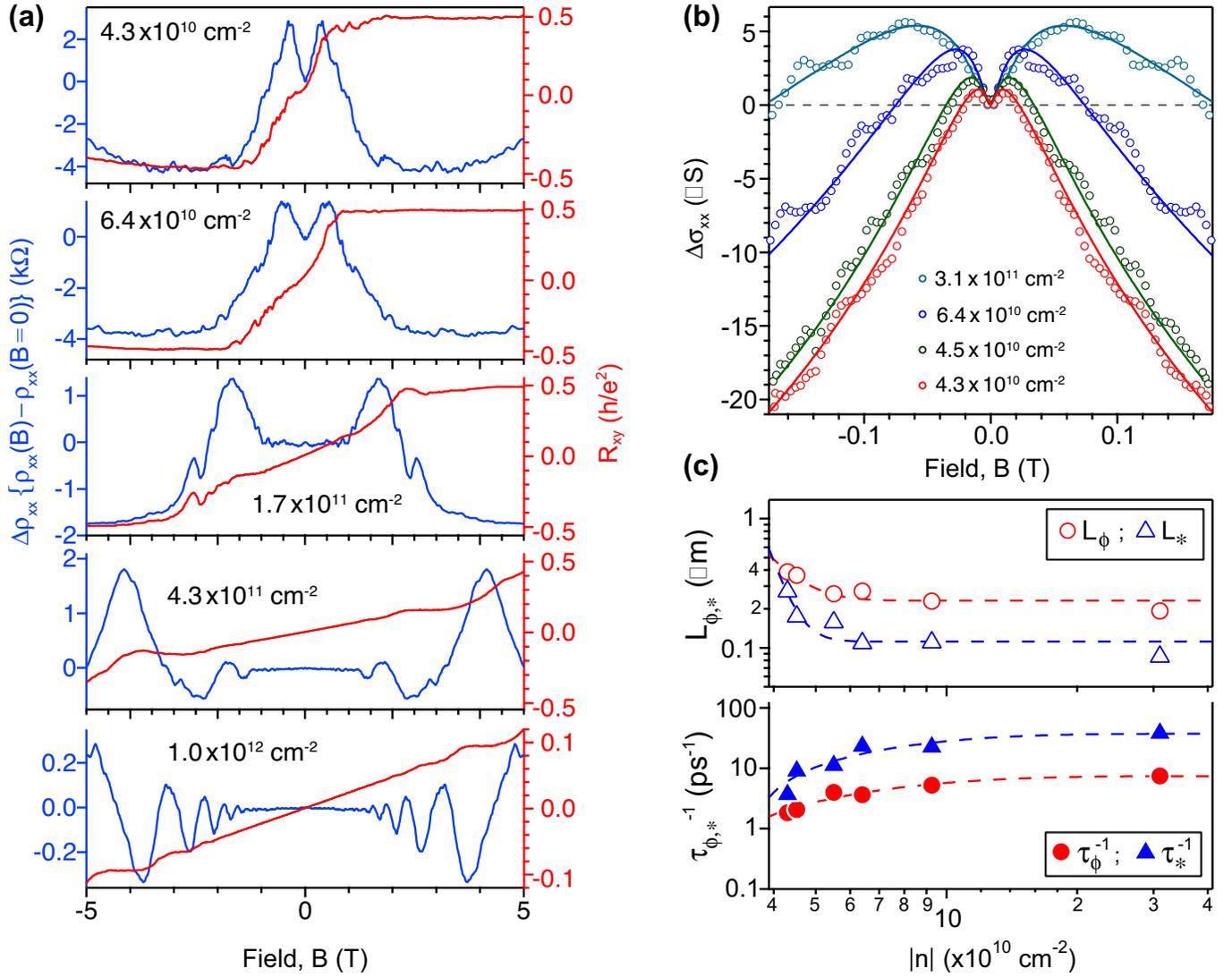

**Figure 4.** Carrier-density-dependent magnetotransport. (a) Magnetoresistance (MR), $\Delta\rho_{xx} = \rho_{xx}(B) - \rho_{xx}(B=0)$ and the Hall resistance $R_{xy}$ data at different carrier densities $n$. At strong-fields ($B \gg B_c \sim 100$ mT, where $B_c$ is the transport field limit) the MR shows well-defined SdH oscillations and corresponding Hall plateaus at each oscillation minimum, characteristic of half-integer quantum Hall effect (QHE). At weak-fields ($B \lesssim B_c \sim 100$ mT), the MR is almost flat at high carrier densities and it becomes positive at low densities as the Fermi level is tuned close to the Dirac point, showing a clear transition from suppressed WL to WAL. (b) Modeling of low-field magnetoconductance (MC) data in the vicinity of the Dirac point, showing a clear transition to WAL below $n \sim 6\times10^{10}$ cm$^{-2}$. (c) Extracted scattering lengths $L_{\phi,*}$ and corresponding scattering rates $\tau_{\phi,*}^{-1}$ obtained from the fits, which show a rapid change as the carrier density $n$ falls below $6\times10^{10}$ cm$^{-2}$. Dashed lines with respective colors correspond to the exponential fits as guide to eyes.

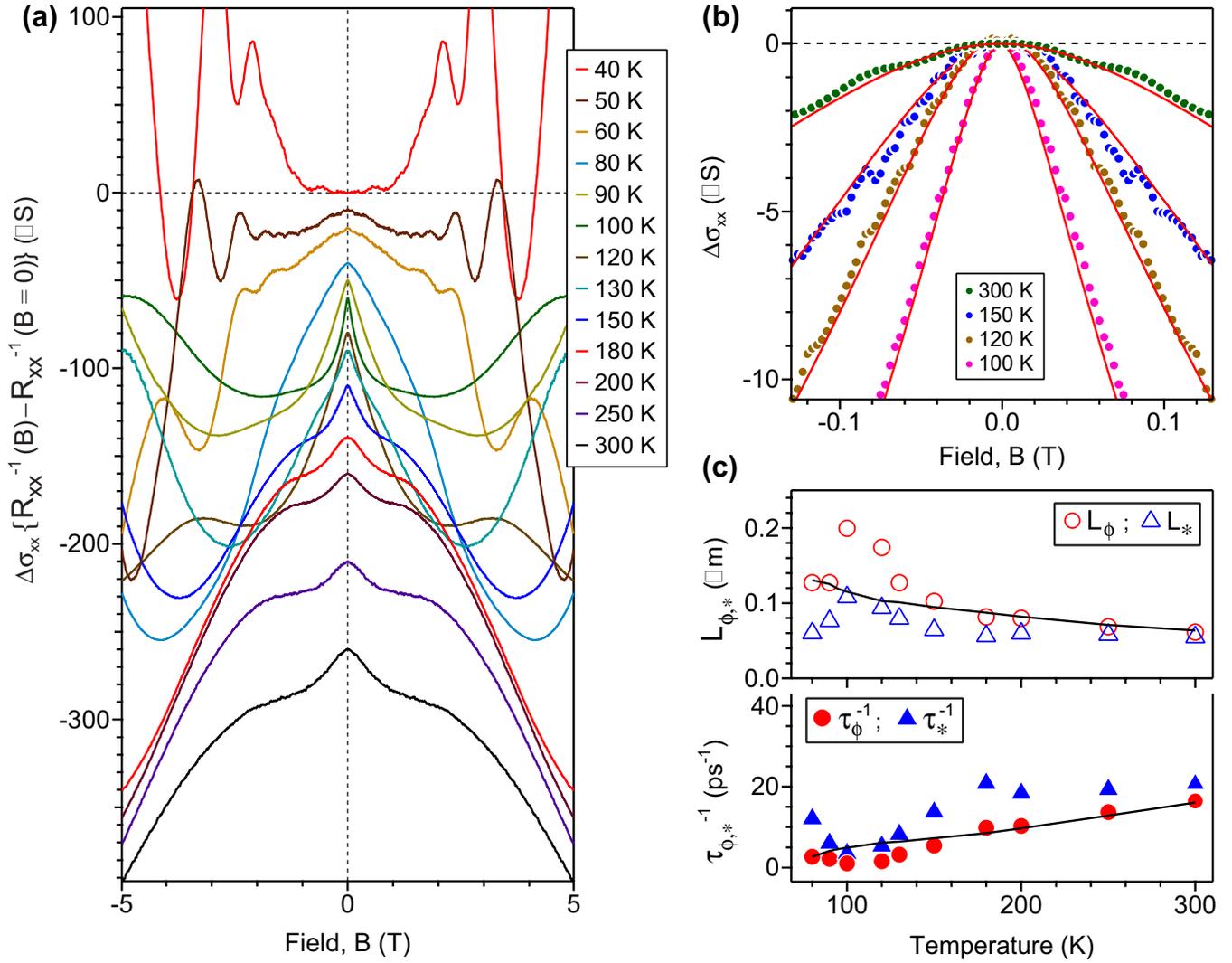

**Figure 5.** Temperature-dependent magnetotransport. (a) Magnetoconductance (MC) data, $\Delta\sigma_{xx} = [R_{xx}^{-1}(B) - R_{xx}^{-1}(B = 0)]$ measured at different temperatures. Each curve (except 40 K data) is shifted equidistance ($-10\ \mu S/10\ K$) vertically for clarity. Below 50 K, the weak-field MC behavior is almost flat while the strong-field behavior is dominated by SdH oscillations. Above 50 K, MC shows a WAL cusp near zero-field embedded on a broad parabolic background. The cusp becomes sharper and stronger at around 100 K and slowly turns broader and weaker beyond 150 K. However, it still appears up to 300 K, showing a clear signature of pseudospin quantum interference up to RT. (b) Weak antilocalization model fittings of the weak-field MC data shown in (a). (c) Extracted scattering lengths $L_{\phi,*}$ and rates $\tau_{\phi,*}^{-1}$ from the fits and plotted as a function of temperature. Black solid

curves are fits to the electron-electron scattering rate $\tau_\phi^{-1}$ and length $L_\phi$ in the ballistic regime, as described by Equation (2).

**Supporting Information:**

# Room-temperature quantum transport signatures in graphene/LaAlO$_3$/SrTiO$_3$ heterostructures

Giriraj Jnawali,[1,2] Mengchen Huang,[1,2] Jen-Feng Hsu,[1,2] Hyungwoo Lee,[3] Jung-Woo Lee,[3] Patrick Irvin,[1,2] Chang-Beom Eom,[3] Brian D'Urso,[1,2] and Jeremy Levy[1,2*]

[1]*Department of Physics and Astronomy, University of Pittsburgh, Pittsburgh 15260, USA*

[2]*Pittsburgh Quantum Institute, Pittsburgh, PA 15260, USA*

[3]*Department of Materials Science and Engineering, University of Wisconsin-Madison, Madison, WI 53706, USA*

*Corresponding Author: Jeremy Levy, (jlevy@pitt.edu)

# SUPPLEMENTARY METHODS

## A. Surface roughness examination

The morphology of graphene/LaAlO$_3$/SrTiO$_3$ (graphene/LAO/STO) and bare LAO/STO surfaces were characterized by atomic force microscope (AFM) in tapping mode. The vertical resolution of our AFM is sub-ångström, which is estimated by resolving the uniform atomic steps on a pristine LAO/STO sample. Lateral resolution is around 5 nm in which the lower limit is defined by the tip radius. A doped silicon (Si) tip with tip radius of ∼5 nm was used for AFM imaging. To avoid any differences caused by the scanning process, identical scanning parameters were used to acquire AFM scans on both surfaces. The AFM images were corrected by third order plane subtraction to compensate for scanning drift. Surface morphology from both surfaces was compared by calculating the standard deviation of the height distribution[1]. Height analysis was performed within a scan size of 2 μm$^2$.

    Figure S 1 shows topographic images and their height histograms obtained from a graphene/LAO/STO sample and a bare LAO/STO surface. Although both surfaces show uniform atomic steps, the LAO/STO region covered by graphene has a reduced root mean square (RMS) roughness compared with the bare LAO/STO surface. Height histograms obtained from both surfaces were fitted using Gaussian distributions and standard deviations $\sigma_{std}$ were determined from best-fit conditions. To avoid any influence from the surface roughness on pristine LAO/STO surface, we have also plotted histograms after masking the regions covered by the atomic steps (see inset of Figure S 1(c)). Both results are summarized in Table S 1. It is to be noted that $\sigma_{std}$ of our graphene/LAO/STO samples (with masking) is appreciably lower than typical $\sigma_{std}$ estimated on good quality graphene/SiO$_2$ samples reported elsewhere[1, 2], confirming improved surface corrugation in our graphene/LAO/STO samples.

## B. Characterization by Raman spectroscopy

Raman measurements were performed on graphene/LAO/STO and bare LAO/STO using a commercial Raman microscope (Renishaw InVia) with 633 nm laser excitation under ambient conditions. The spectral resolution of the instrument was approximately 1 cm$^{-1}$. We used a 50× microscope objective (numerical aperture $NA = 0.73$) and laser power of ~7 mW. Slightly higher laser power was necessary to collect better spectra because of the transparent substrate. Raman spectra were acquired at different positions on the sample and several acquisition cycles were repeated at each spot to check possible damage caused by laser-induced heating effects. No drastic change in Raman intensity of prominent peaks confirmed negligible heating effect in our samples. Surface morphology was re-examined around the laser exposure region by using non-contact AFM scanning and found no evidence of surface degradation.

Typical Raman spectra, which are consistently reproduced at each region on the sample, are presented in Figure S 2. On the graphene/LAO/STO regions, sharp G- and 2D-peaks are visible in addition to broad peaks at 1320 cm$^{-1}$ and 1610 cm$^{-1}$. Those broad peaks are also reproduced in the spectrum recorded on bare LAO/STO, indicating that their origin is purely coming from LAO/STO. Those peaks are indeed second-order Raman peaks of STO, as expected from its cubic perovskite structure[3]. To separate the Raman peaks arising only from the graphene layer (net spectrum), the spectrum recorded on bare LAO/STO is subtracted from the spectrum recorded on graphene/LAO/STO, assuming no interaction between them (black solid line in Figure S 2). The net spectrum displays prominent phonon modes of graphene, i.e., G- and 2D-modes. The absence of a D-mode peak at 1350 cm$^{-1}$ indicates very low defect density in our samples. Both G-mode and 2D-mode peak positions and their widths are analyzed by fitting with a Lorentzian function. G-mode peak appears at around $1586 \pm 1$ cm$^{-1}$ with a FWHM = $7.5 \pm 1$ cm$^{-1}$ and 2D-mode peak appears at $2645 \pm 1$ cm$^{-1}$ with a FWHM = $24.5 \pm 1$ cm$^{-1}$. The error bars are estimated by averaging different spectra acquired from different positions on the sample. The intensity ratio between G-mode and 2D-mode peaks is estimated to be $I_{2D}/I_G \sim 3.8$ and is even higher for some samples. The G-mode peak is upshifted by $\sim 6\,\text{cm}^{-1}$ and 2D-mode is down shifted by $\sim 35\,\text{cm}^{-1}$ as compared to the Raman modes from undoped pristine graphene; an opposite shift of the G and 2D peaks corresponds to electron doping of the graphene[4, 5]. Since the ratio between G and 2D peak intensities is relatively larger than typical CVD-grown graphene on $SiO_2$ samples, we can qualitatively estimate very low doping density. In addition, the intensity variation is always $I_{2D}/I_G > 3$ and the 2D-mode shows a single Lorentzian profile, providing confirmation of single layer graphene in our samples.

We also observe an asymmetric line shape of the G-mode peak in the subtracted spectrum. The observation of a Fano-like asymmetric line shape could be due to phonon renormalization via coupling with photoinduced rapid polarization fluctuations in underlying substrate. This observation also provides interesting prospects for future investigations in this system.

## C. Graphene patterning into Hall bar structure via *in situ* c-AFM lithography

We used an AFM system (Asylum Research MFP-3D) with an environmental closure with controlled humidity ($\sim 30\,\%$) for graphene patterning process. A conductive Si tip was used for both imaging and lithography. Imaging was performed in non-contact (tapping) mode and lithography was performed in contact mode. Figure S 3(a) demonstrates typical patterning process, showing how graphene layer is etched out or cut and electrically isolate by using a c-AFM lithography. A negative bias voltage (amplitude $\sim -25\,\text{V}$) was applied to the tip to sufficiently locally etch out the graphene surface. Graphene conductance (between the electrodes) is monitored during lithography process, which provides confirmation of electrical isolation once patterning was done across the graphene.

Figure S 3(b) shows a test of lithography process using a graphene/LAO/STO sample, where graphene stripes are patterned on LAO/STO connecting with gold electrodes. As soon as graphene is detached or etched away, the conductance drops sharply as shown by the arrow in the plot. This method was applied to pattern several Hall bar devices as shown in Figure S 4(a)-(c). Since this method is automated to create a variety of shapes such as lines, rectangles, circles, and also a single dot via pulsed voltage, it is very useful to fabricate various types of nanostructures or quantum devices without undergoing additional processing steps such as electron-beam lithography and/or etching.

## D. Characterization of resistance hysteresis

Electric field gating of graphene devices through $LaAlO_3/SrTiO_3$ interface produces unusual ferroelectric like resistance hysteresis in which the Dirac point shifts after reversal of the gate sweep direction. The voltage hysteresis at the Dirac point varies with sweeping parameters such as sweep range and rate. In general, the hysteresis increases as the sweeping range and the rate increase. At a fixed sweep rate and sweep range, the hysteresis stays unchanged and is reproduced consistently during the measurements. A similar hysteresis effect was previously observed in graphene devices on ferroelectric substrates[6] and was attributed to the response of ferroelectric polarization switching. Graphene devices on $SrTiO_3$ also exhibited similar hysteresis behavior and it was argued that the surface dipole moment associated with a puckered oxygen layer on the $SrTiO_3$ surface was responsible for the hysteresis[7, 8]. Since our graphene devices are fabricated on top of $LaAlO_3$, both scenarios are unlikely to apply to our samples. Direct electrochemical doping due to charged polar molecules (e.g., $H_2O$ and $O_2$) on graphene[9, 10] also has a minimal effect in our devices because our samples are annealed electrically by passing large currents (~150 nA) for several hours at 350 K under moderate vacuum (1 − 10 Torr) before performing transport measurements. In addition, AFM analysis shows a clean graphene surface without any layers or clusters of water molecules (see Figure S 1 and Figure S 4). More evidence comes from Raman data and Hall measurements at RT, which show slight electron doping in our samples in contrast to the expected hole doping from adsorbed water molecules. Therefore, based on experimental observations, it is reasonable to associate the observed hysteresis behavior in graphene with previously reported field-induced hysteretic conductance of critical-thickness $LaAlO_3/SrTiO_3$ interfaces[11]. A clearer understanding of the origin of the observed hysteresis will require more detailed investigations, and is beyond the scope of the work presented here.

# SUPPLEMENTARY TABLE

**Table S 1.** Surface roughness examination by height histogram. Standard deviation and width of the height histogram measured on bare LAO/STO and graphene/LAO/STO surfaces.

|  | LAO/STO (with steps) | Graphene/LAO/STO (with steps) | LAO/STO (no steps) | Graphene/LAO/STO (no steps) |
|---|---|---|---|---|
| $\sigma_{std}$ | 184 pm | 162 pm | 168 pm | 134 pm |
| $\Gamma_{width}$ | 0.433 nm | 0.381 nm | 0.39 nm | 0.31 nm |

# SUPPLEMENTARY FIGURES

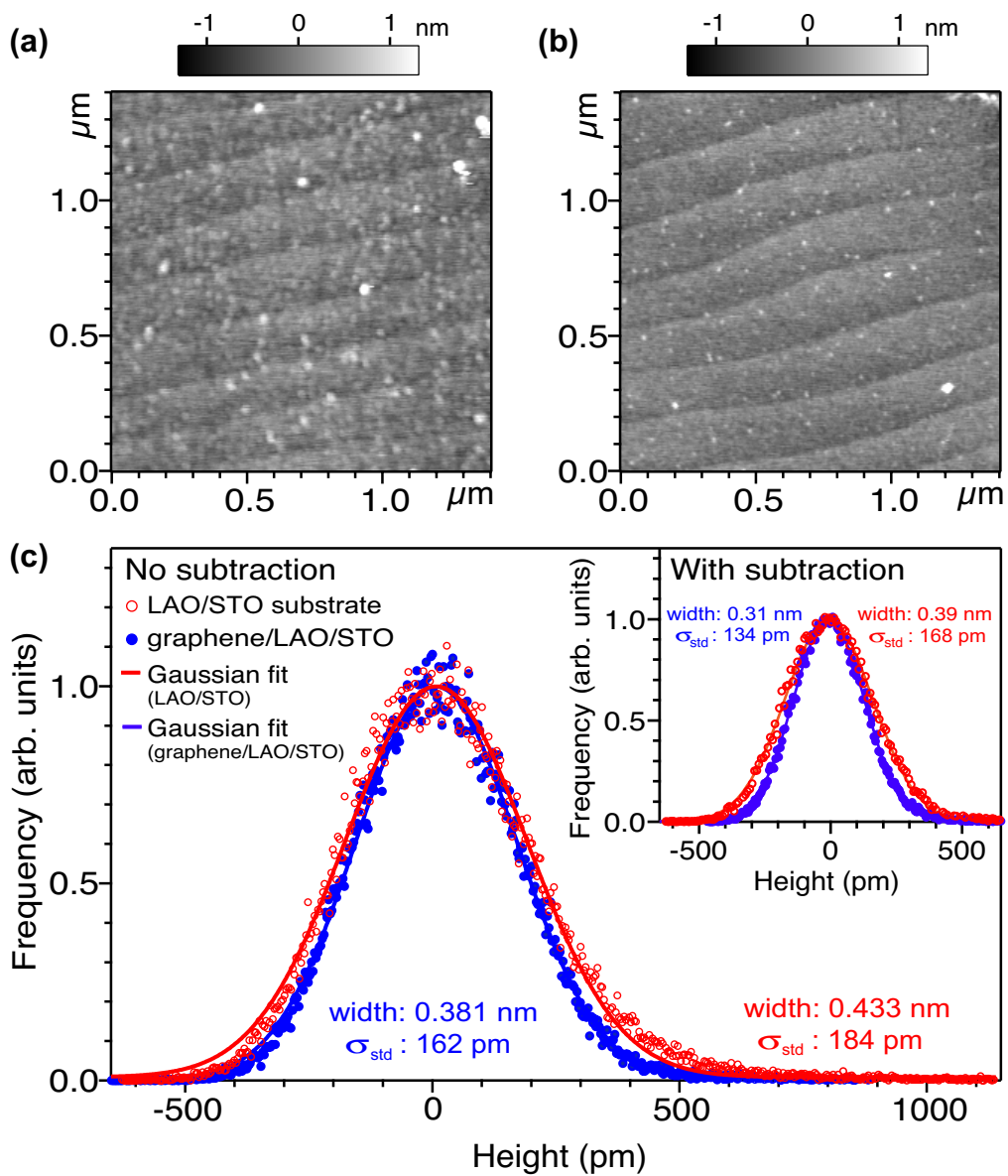

**Figure S 1.** Comparison of surface morphology of bare LAO/STO and graphene/LAO/STO. Tapping mode AFM height images of (a) bare LAO/STO and (b) graphene/LAO/STO surfaces. Both images are presented with the same height scale. (c) Height histograms of the data in (a) as red open circles and in (b) as blue filled circles. The histograms are well described by Gaussian distributions, as shown by red and blue solid lines for each plot. Inset shows the respective histogram data after masking the regions covered by atomic steps of underlying LAO/STO substrate.

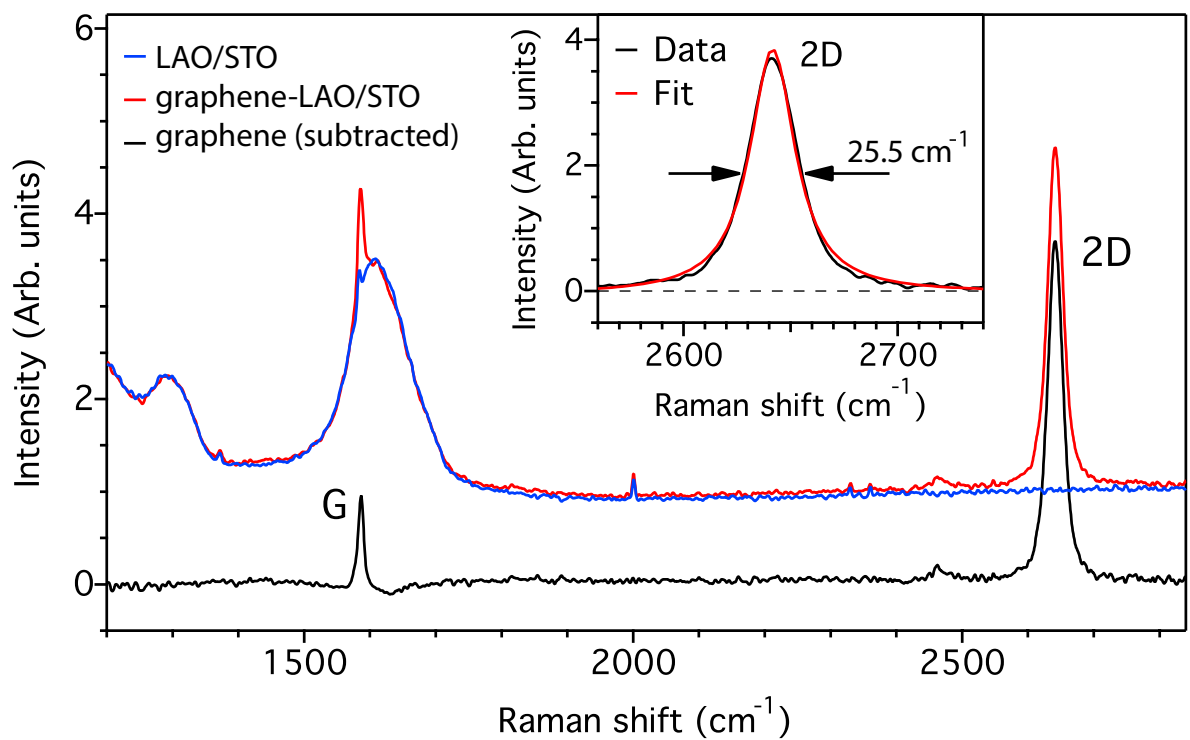

**Figure S 2.** Raman characterization of graphene/LAO/STO sample**.** Raman spectra measured on graphene/LAO/STO and bare LAO/STO regions on the sample surface for an excitation wavelength of 633 nm. Sharp peak at 1586 cm$^{-1}$ and a single Lorentzian peak at 2645 cm$^{-1}$ on graphene/LAO/STO region correspond to the typical G-mode and 2D-mode optical phonon in graphene. Additional two broad peaks at around 1320 cm$^{-1}$ and 1610 cm$^{-1}$, which are also present on bare LAO/STO region, correspond to the second-order Raman modes of STO. Net graphene spectrum, extracted by subtracting two spectra, is also shown with solid black line. It is clear from the subtracted spectrum that there is no measurable D peak, indicating low defect density in graphene.

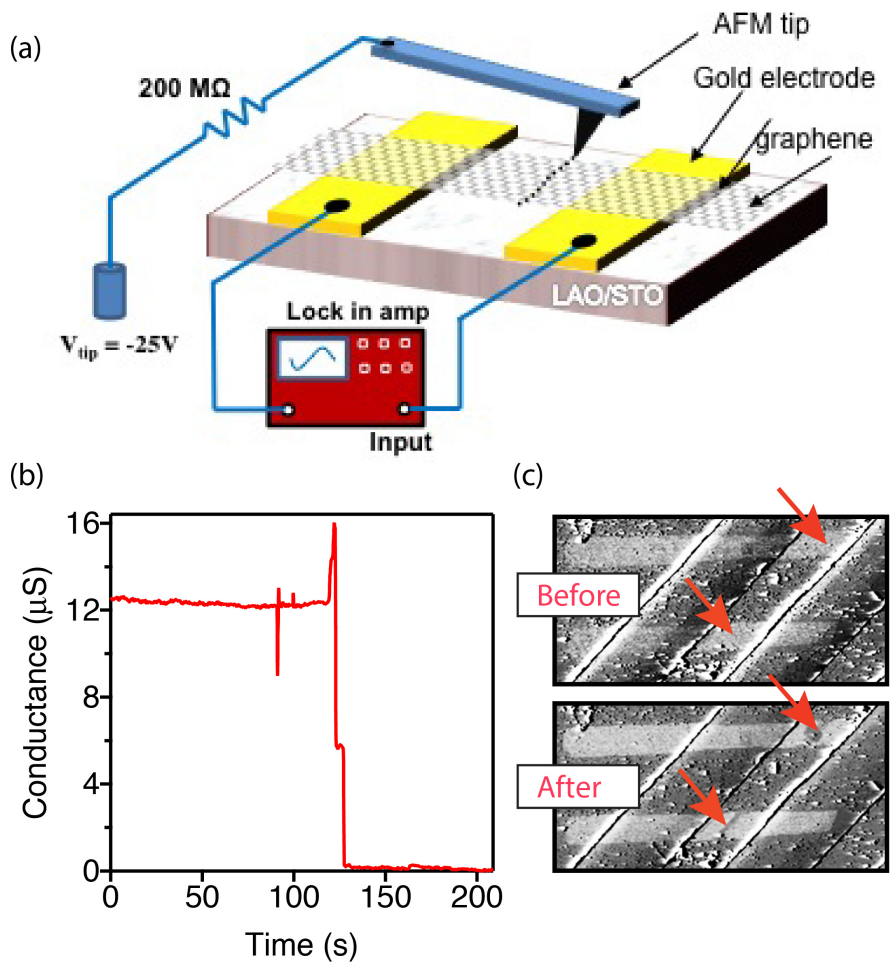

**Figure S 3.** Graphene patterning by c-AFM lithography. (a) Schematic view of a test sample for c-AFM lithography in which graphene is contacted by top-gated gold electrodes on the LAO/STO heterostructure. (b) Negative DC-voltage is applied to the AFM tip and it is scanned across the graphene stripe in a contact mode under a controlled conditions such as scanning speed (300 nm/s) and resistance between the electrode is monitored by using lock-in amplifier. (b) and (c) Experimental demonstration of graphene patterning by c-AFM lithography on a graphene/LAO/STO sample. As soon as graphene is etched out, conductance drops to zero, indicating complete breaking of electrical path between the electrodes. Tapping mode AFM images acquired before and after the lithography process show complete breaking of graphene along the written path, as indicated by the red arrows.

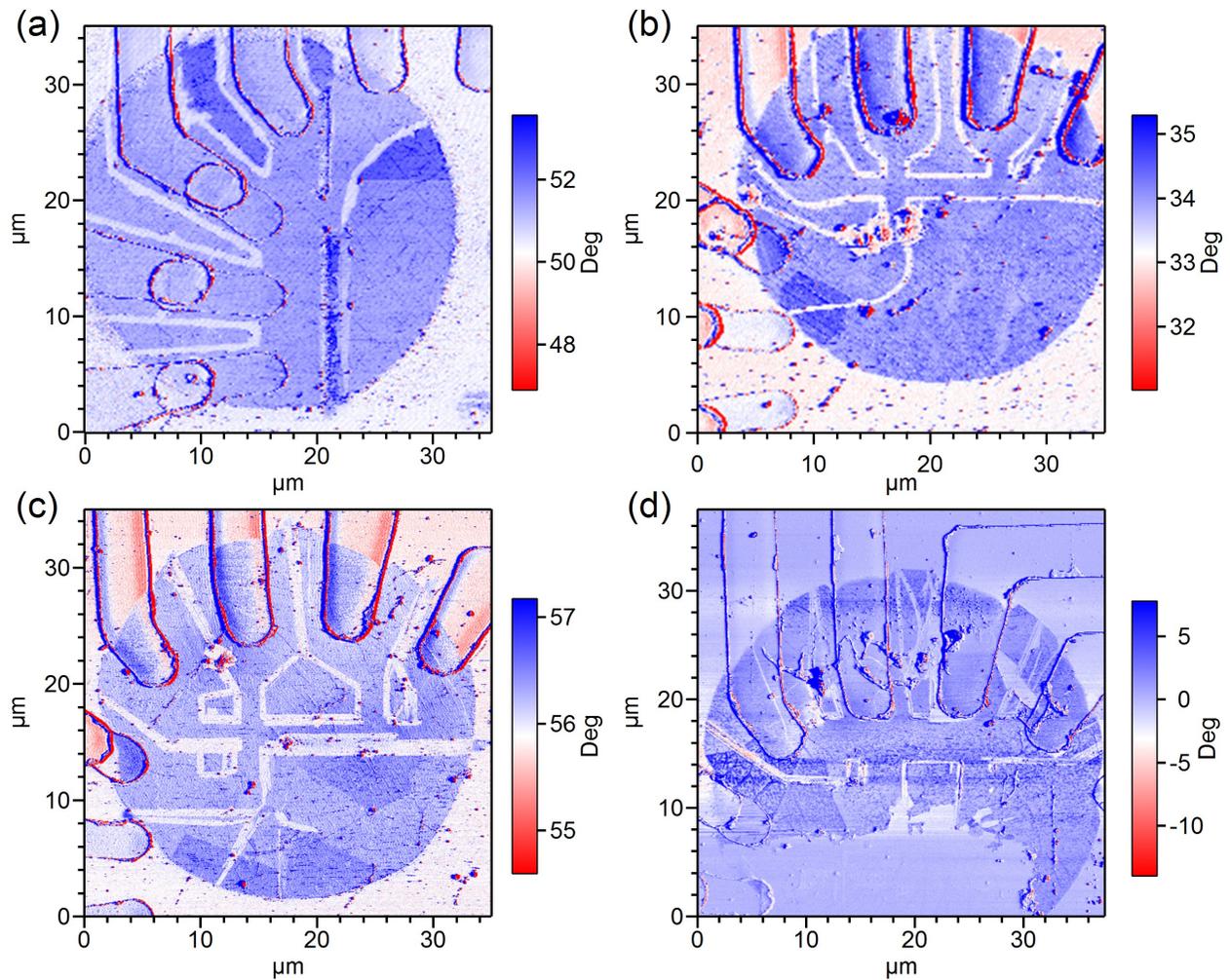

**Figure S 4.** AFM patterned Hall bar devices on graphene/LAO/STO. (a) – (d) Tapping mode AFM phase images of patterned graphene/LAO/STO hall bars of different widths and lengths. Phase images are shown because of high contrast between graphene/LAO/STO and bare LAO/STO surface. Brighter region is defined for LAO/STO surface where graphene is etched out by using c-AFM lithography. We can also recognize few layers and multilayer graphene in some regions (dark blue regions), which is otherwise difficult to recognize in height images. Device (c) has multilayer graphene directly on the device channel, demonstrating that this technique is equally applicable for multilayer graphene devices. The channel width of the device can be optimized by varying various parameters such as sharpness of the tip (i.e., tip radius), tip voltage, and scanning speed, which are not shown in this work.

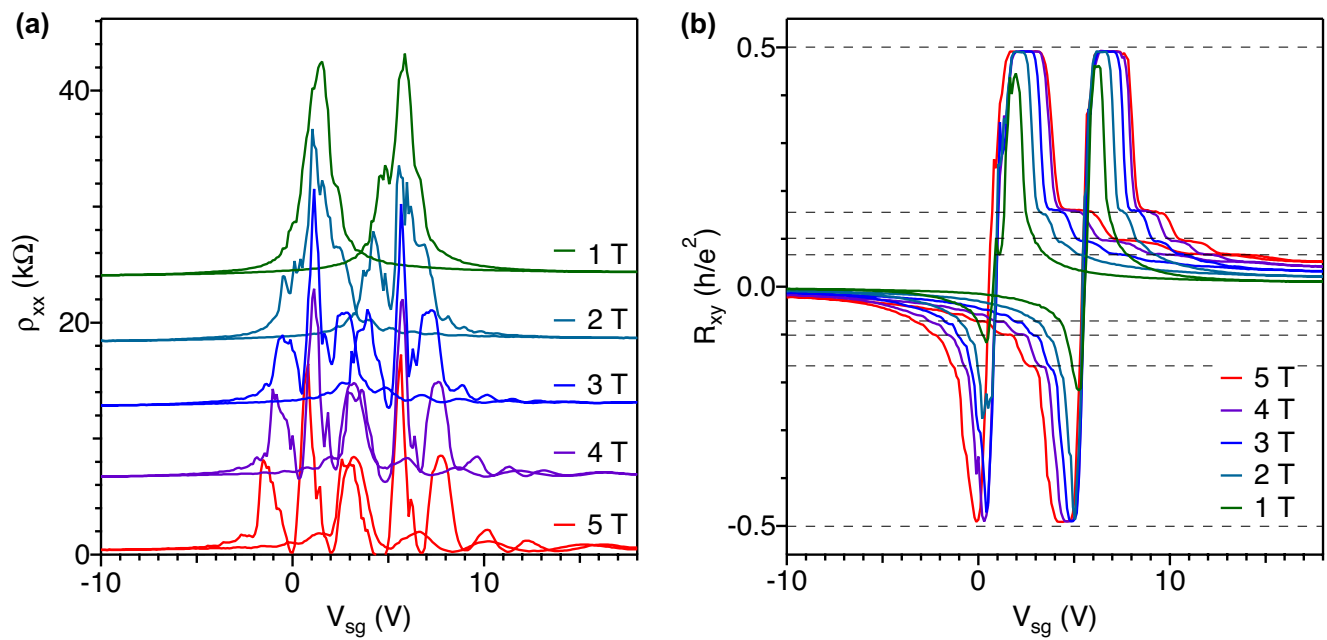

**Figure S 5.** Field-dependent side gate tuning of graphene resistivity. (a) Side gate ($V_{sg}$) dependence of graphene resistivity at different magnetic fields applied normal to the graphene surface. Hysteresis appears when gating is reversed. Pronounced SdH oscillations develop as the field is increased and the oscillations are equally visible at both sides of hysteresis. (b) Simultaneously measured side gate dependence of Hall resistance at different fields. Hall plateaus develop at each minimum of SdH oscillations at high fields, showing QHE equally appears at both sides of hysteresis.

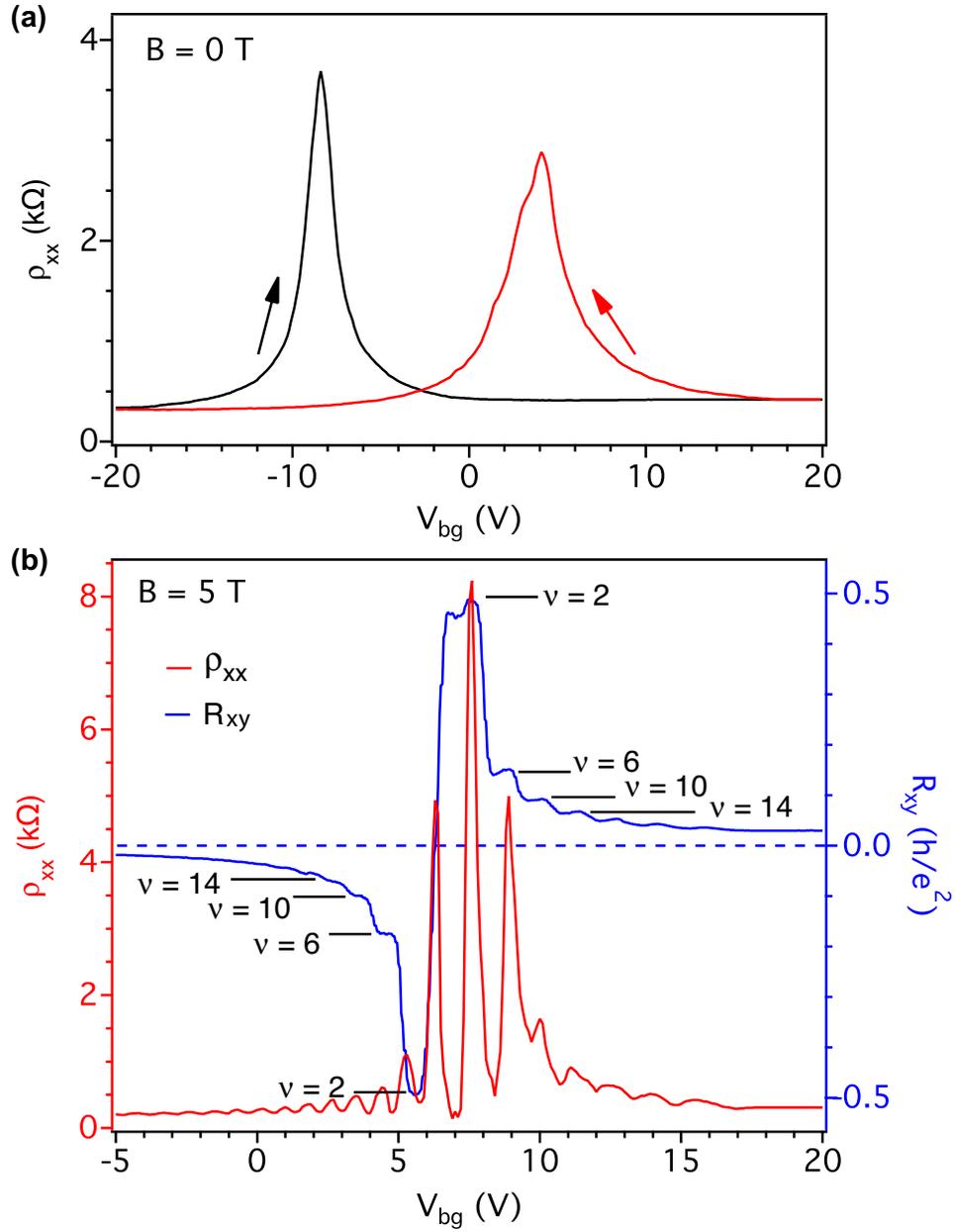

**Figure S 6.** Field-effect tuning of graphene resistivity using back gate through bottom of the SrTiO$_3$ substrate. (a) Back gate dependence of longitudinal resistivity $\rho_{xx}$ measured at zero field $B = 0$ T and at 2 K. Hysteresis is observed when the gate tuning direction is reversed. (b) Simultaneously measured back gate dependence of longitudinal resistivity $\rho_{xx}$ and Hall resistance $R_{xy}$ measured at fixed field of $B = 5$ T and at 2 K. Since all features are identical at each side of hysteresis only the right side of gate tuning is shown.

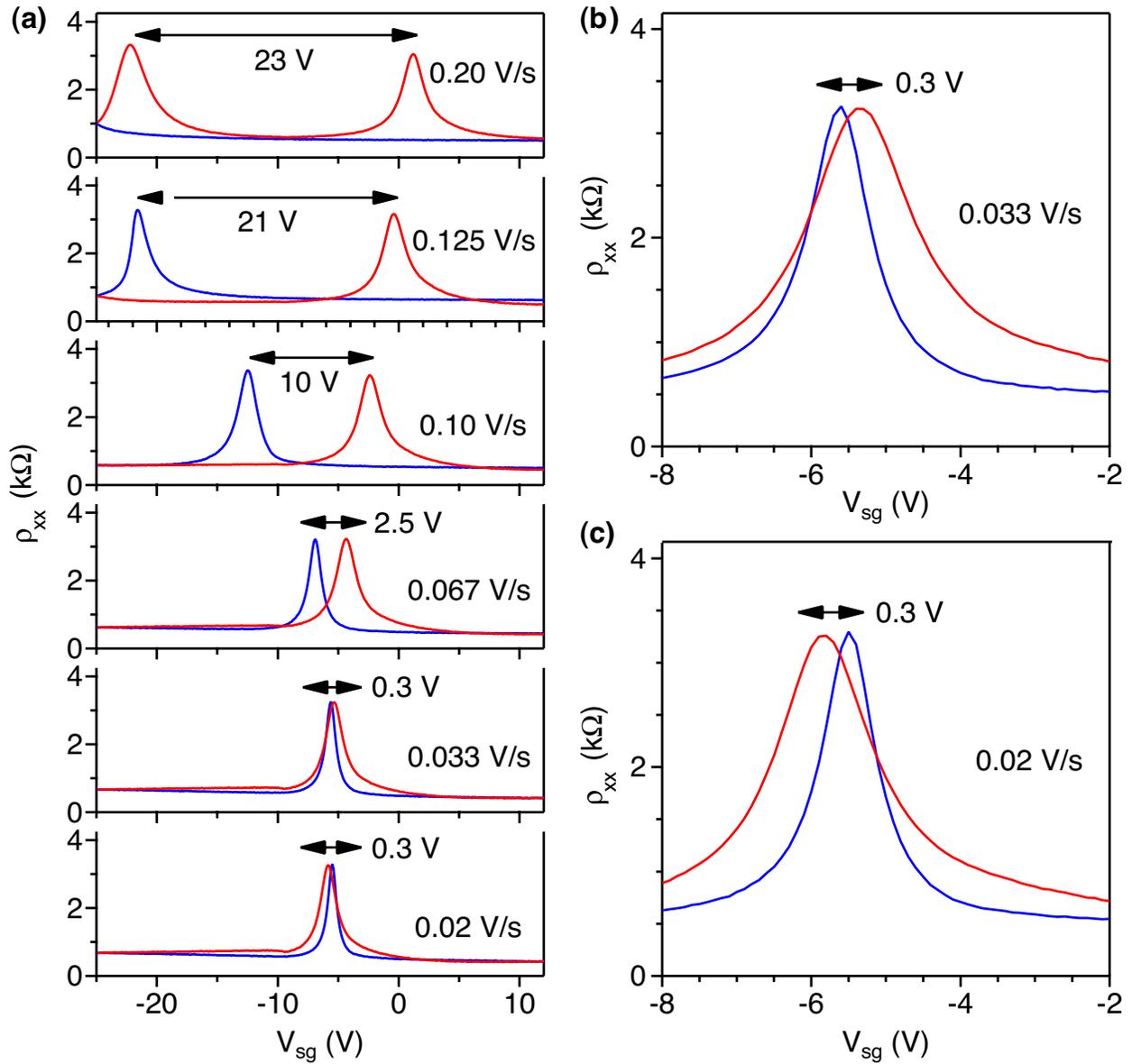

**Figure S 7.** Hysteresis in the resistivity of a graphene/LAO/STO device during the gate voltage sweep. (a) Change of Dirac point shift as a function of sweep rate measured at 350 K. As the sweep rate is decreased, the Dirac point shift decreases significantly and even goes to negative shift. (b) and (c) Close-up plots for two lowest sweep rates, i.e., 0.033 V/s and 0.02 V/s, respectively. Negative shift of dual Dirac points can be identified from their asymmetric shape as they shift from positive to negative.

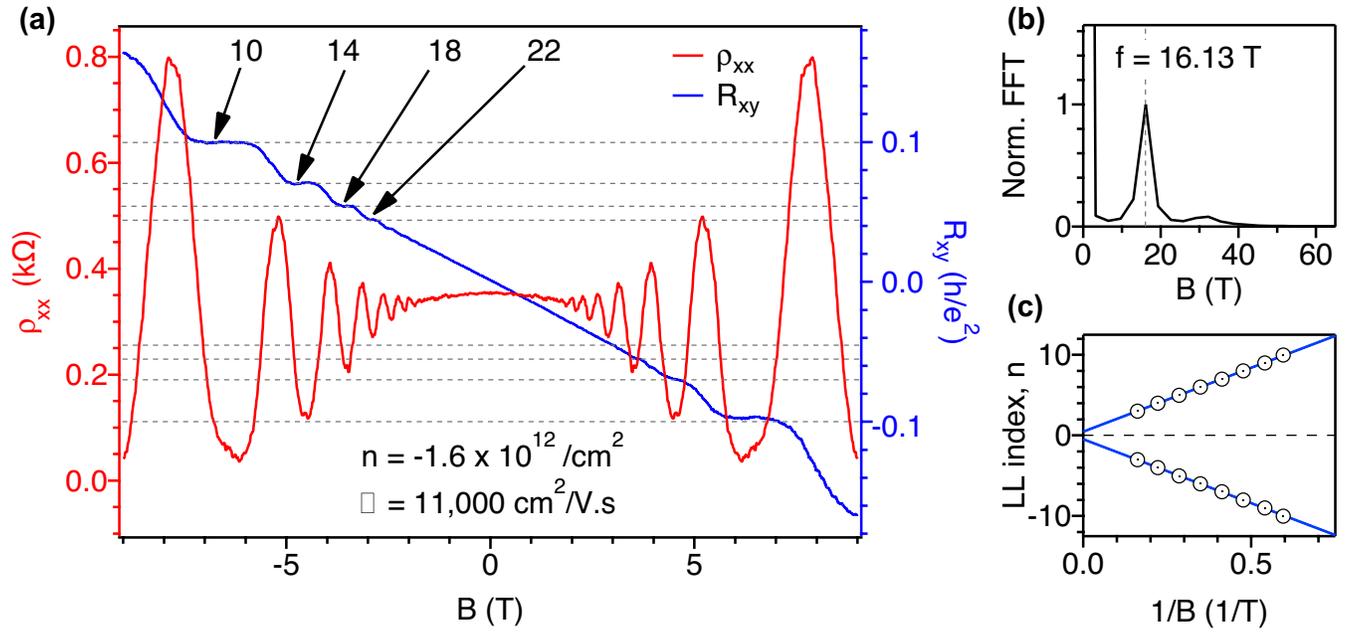

**Figure S 8.** Magnetotransport of a graphene/LAO/STO device and Berry's phase. (a) The sheet resistivity $\rho_{xx}$ and the Hall resistance $R_{xy}$ as a function of out-of-plane magnetic field up to $B = \pm 9$ T at $V_{sg} = 0$, measured at 0.1 K. Pronounced SdH oscillations and quantized Hall plateaus at the minima of the oscillations are indicated by different Landau filling factors $\nu = 10, 14, 18,$ and $22$. (b) Frequency of the SdH oscillations $f = 16.13$ T is extracted for the fast Fourier transform (FFT) of the $\rho_{xx}$ vs $1/B$ plot. (c) Plot of Landau level (LL) index as a function of $1/B$. The solid lines correspond to a linear fit with a slope of 16.13 T, in which the intercept of 0.45 estimates a Berry's phase of $\pi$ in graphene.

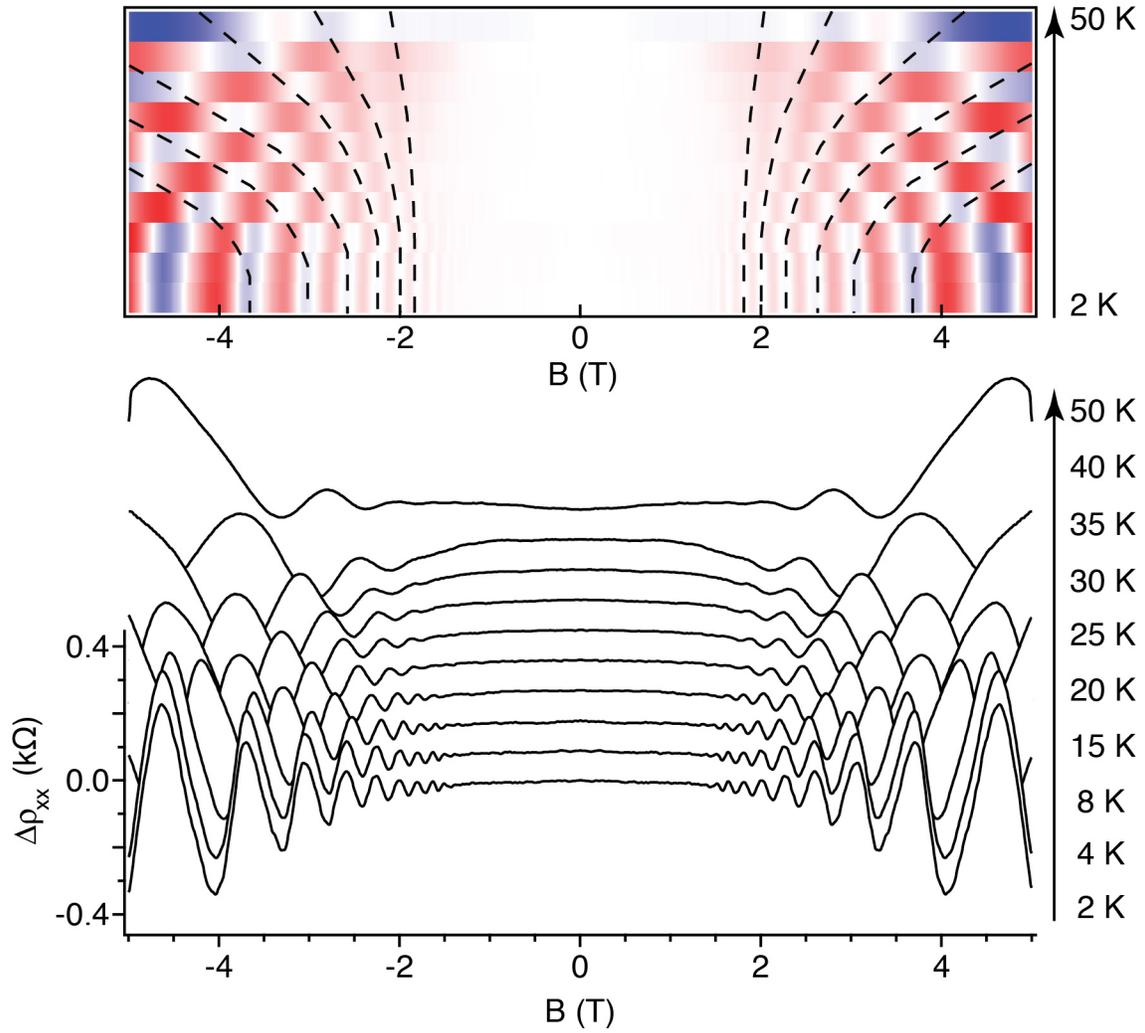

**Figure S 9.** Magnetic field sweep at different temperatures. Sheet resistivity $\rho_{xx}$ of a graphene device measured during sweeping of out-of-plane magnetic field from $+5$ T to $-5$ T at different temperatures. Each plot shows pronounced Shubnikov-de Haas (SdH) oscillations after $\pm 1$ T and the oscillation amplitude gradually decreases as the sample is warmed up. The decrease of amplitude is caused by thermal broadening of LLs. In addition to the amplitude, the oscillation period also gradually increases with increasing the sample temperature. For better visibility, a 2D color plot is also shown on top where black dashed lines indicate a clear shift of oscillation peaks, suggesting a gradual change of the Fermi level towards the Dirac point as the sample temperature increases.

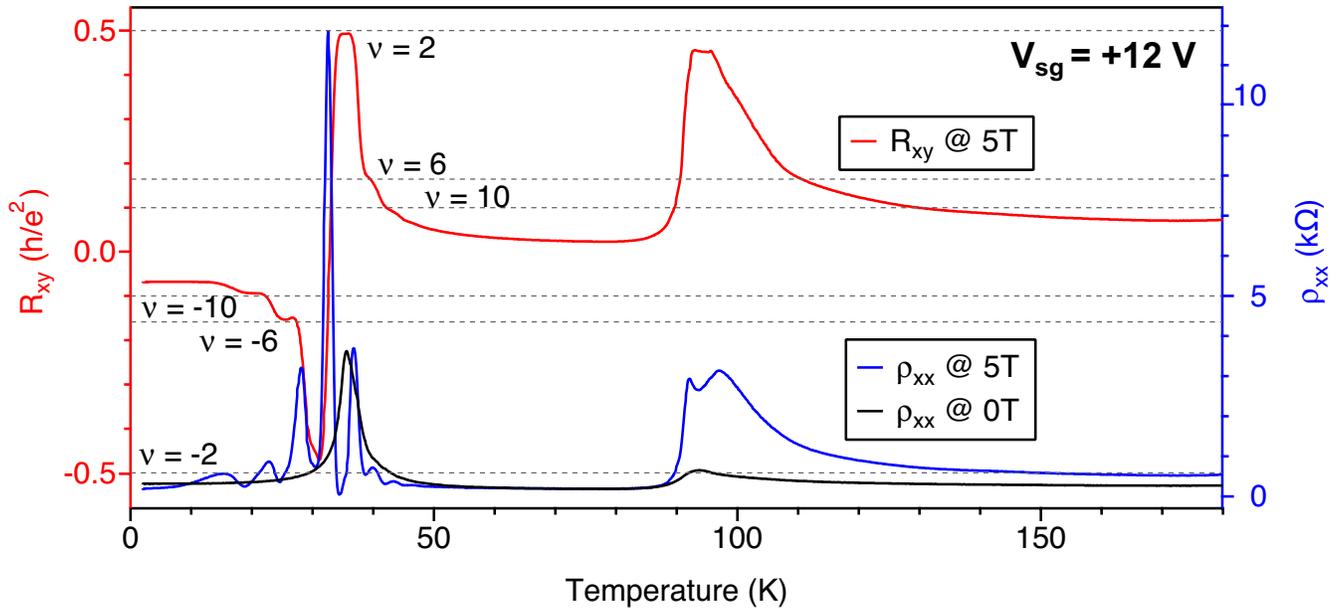

**Figure S 10.** Temperature dependence of magnetoresistance with a fixed side gate. Graphene sheet resistivity $\rho_{xx}$ and Hall resistance $R_{xy}$ as a function of temperature at a fixed magnetic field of $B = 0$ T and 5 T, measured with a fixed side gate ($V_{sg} = +12$ V). At zero field $B = 0$ T, sheet resistivity $\rho_{xx}$ peaks at around 35 K and 90 K, which is in contrast to a single broad peak around 90 K without gating the device. With an applied field of $B = 5$ T, sheet resistivity $\rho_{xx}$ shows pronounced Shubnikov-de Haas (SdH) oscillations and well-defined half-integer plateaus of the Hall resistance as the sample is warmed up from 2 K to room temperature. The Hall plateaus match perfectly with theoretically predicted Landau levels in single layer graphene with filling factors $\nu = \pm 2, \pm 6, \pm 10, ..$, demonstrating a clear indication of temperature-dependent Hall quantization. Origin of broad peak at around 90 K is most likely caused by temperature dependence of structural phase transition in $SrTiO_3$. However, peak at 35 K may be caused by spontaneous non-linear response of carrier tuning presumably associated with interactions with the substrate.

# Supplementary References

[1]     C. H. Lui, L. Liu, K. F. Mak, G. W. Flynn, T. F. Heinz, Nature 2009, 462, 339.

[2]     M. Ishigami, J. H. Chen, W. G. Cullen, M. S. Fuhrer, E. D. Williams, Nano Letters 2007, 7, 1643.

[3]     W. G. Nilsen, J. G. Skinner, The Journal of Chemical Physics 1968, 48, 2240.

[4]     J. Yan, Y. Zhang, P. Kim, A. Pinczuk, Phys. Rev. Lett. 2007, 98, 166802.

[5]     M. Kalbac, A. Reina-Cecco, H. Farhat, J. Kong, L. Kavan, M. S. Dresselhaus, ACS Nano 2010, 4, 6055.

[6]     X. Hong, J. Hoffman, A. Posadas, K. Zou, C. H. Ahn, J. Zhu, Appl. Phys. Lett. 2010, 97, 033114.

[7]     S. Saha, O. Kahya, M. Jaiswal, A. Srivastava, A. Annadi, J. Balakrishnan, A. Pachoud, C. T. Toh, B. H. Hong, J. H. Ahn, T. Venkatesan, B. Ozyilmaz, Sci. Rep. 2014, 4, 6173.

[8]     R. Sachs, Z. S. Lin, J. Shi, Sci. Rep. 2014, 4, 3657.

[9]     H. M. Wang, Y. H. Wu, C. X. Cong, J. Z. Shang, T. Yu, Acs Nano 2010, 4, 7221.

[10]    H. Xu, Y. Chen, J. Zhang, H. Zhang, Small 2012, 8, 2833.

[11]    S. Thiel, G. Hammerl, A. Schmehl, C. W. Schneider, J. Mannhart, Science 2006, 313, 1942.